\documentclass{article}
\usepackage{iclr2025_conference,times}

\usepackage[colorlinks=true,citecolor=brown,urlcolor=gray]{hyperref}
\usepackage[skip=3pt]{caption}

\usepackage{tcolorbox}
\usepackage{pgfplots,pgfplotstable}
\usepgfplotslibrary{groupplots}
\usepackage{pifont}

\newcommand{\cmark}{\textcolor{teal}{\ding{51}}} 
\newcommand{\xmark}{\textcolor{red}{\ding{55}}} 

\newcommand{\modelname}{\textcolor{purple}{\textit{A\textsuperscript{2}SB}}\xspace}
\usepackage{bbm} 

\usepackage{siunitx}
\usepackage{colortbl}
\newcommand{\bestnumber}[1]{\cellcolor{pink}{#1}}

\usepackage{amsmath,amsfonts}
\usepackage{microtype}
\usepackage{graphicx}
\usepackage{subcaption}
\usepackage{booktabs} 

\PassOptionsToPackage{table}{xcolor}
\usepackage{pgfplots,pgfplotstable}
\usetikzlibrary{pgfplots.statistics, arrows.meta, positioning, calc}
\pgfplotsset{compat=1.18}
\usepackage{xcolor}

\usepackage[T1]{fontenc}      
\usepackage{lmodern}          

\usepackage{multirow}
\usepackage{amssymb}

\usepackage{hyperref}
\usepackage{url}

\usepackage[capitalize,noabbrev]{cleveref}

\definecolor{forestgreen}{rgb}{0.13, 0.55, 0.13}
\definecolor{fulvous}{rgb}{0.86, 0.52, 0.0}
\definecolor{glaucous}{rgb}{0.38, 0.51, 0.71}
\definecolor{lava}{rgb}{0.81, 0.06, 0.13}
\definecolor{buff}{rgb}{0.94, 0.86, 0.51}
\definecolor{chromeyellow}{rgb}{1.0, 0.65, 0.0}
\definecolor{brightube}{rgb}{0.82, 0.62, 0.91}

\definecolor{CustomBlue}{RGB}{76, 114, 176}
\definecolor{CustomRed}{RGB}{196, 78, 82}

\hypersetup{
    pdfborderstyle={/S/U/W 1}, 
    linkbordercolor=red,       
    citebordercolor=green,     
    filebordercolor=magenta,   
   urlbordercolor=cyan        
}

\pgfplotsset{scaled y ticks=false, scaled x ticks=false}

\usepackage{interval}
\usepackage{tabularx}
\usepackage{color}
\usepackage{xspace}
\usepackage{mathtools}  
\usepackage{siunitx}    

\usepackage{placeins}
\usepackage{algorithm}

\newcommand{\argmin}{\operatorname*{argmin}}

\newcommand{\Sch}{Schr\"{o}dinger}
\newcommand{\pdata}{p_{\mathrm{data}}}

\usepackage[algo2e,ruled,vlined,linesnumbered]{algorithm2e} 

\title{
\centering
\selectfont\fontfamily{pzc}\selectfont
\modelname: Audio-to-Audio Schr\"odinger Bridges
}
\author{
    Zhifeng Kong\thanks{Equal contribution.},
    Kevin J Shih\footnotemark[1],
    Weili Nie,
    Arash Vahdat,
    Sang-gil Lee,
    \\
    \textbf{
    João Felipe Santos,
    Ante Jukić,
    Rafael Valle,
    Bryan Catanzaro}\\
    \newline 
    \\
    ~\textbf{NVIDIA} \\ \vspace{2mm}
    ~\texttt{\{zkong,kshih\}@nvidia.com}
}

\iclrfinalcopy

\begin{document}

\maketitle
\begin{abstract}
Real-world audio is often degraded by numerous factors. This work presents an audio restoration model tailored for high-res music at 44.1kHz. Our model, Audio-to-Audio \Sch{} Bridges (\modelname), is capable of both bandwidth extension (predicting high-frequency components) and inpainting (re-generating missing segments). Critically, \modelname is end-to-end requiring no vocoder to predict waveform outputs, able to restore hour-long audio inputs, and trained on permissively licensed music data. \modelname is capable of achieving state-of-the-art bandwidth extension and inpainting quality on several out-of-distribution music test sets. 
~~\newline
~~\newline
{
\footnotesize
\textbf{GitHub Page:} \url{https://github.com/NVIDIA/diffusion-audio-restoration}

\textbf{Checkpoints:} \url{https://huggingface.co/nvidia/audio_to_audio_schrodinger_bridge}

\textbf{Demo Website:} \url{https://research.nvidia.com/labs/adlr/A2SB/}
}

\end{abstract}

\section{Introduction}

Audio signals in the real world may be perturbed due to numerous factors such as recording devices, data compression, and online transferring. For instance, certain recording devices and compression methods may result in low sampling rate, and online transferring may cause a short audio segment to be lost. These problems are usually ill-posed \citep{narayanaswamy2021design, moliner2023solving} and are usually solved with data-driven generative models. For instance, bandwidth extension methods have been proposed to up-sample the audio \citep{lee2021nu,liu2022neural,serra2022universal,moliner2022behm,shuai2023mdctgan,yu2023conditioning,kim2024audio,liu2024audiosr,ku2024gen,yun2025flowhigh}, and inpainting methods have been developed to predict segments where audio is missing \citep{marafioti2019context,marafioti2020gacela,borsos2022speechpainter,liu2023maid,moliner2023diffusion,asaad2024fill}.

Many of these methods are task-specific, designed for the speech domain, or trained to only restore the degraded magnitude -- which requires an additional vocoder to transform the restored magnitude into a waveform. In this work, we investigate high-res music restoration, a more challenging task than speech restoration in terms of typical bandwidth of speech ($\leq$22.05kHz) vs music signals (44.1kHz). We aim to tackle bandwidth extension and inpainting in a single model as in practice it is easier to maintain one model serving multiple tasks. We also aim to build an end-to-end trainable generative model for audio restoration without need of a separate vocoder or a codec. To achieve our goals, we adopt the \Sch{} Bridge framework \citep{de2021diffusion,chen2021likelihood,liuI2SB,albergo2023stochastic} as it is suitable for translation tasks where a part of the source and target samples are well aligned. We name our model \textcolor{purple}{\textit{\modelname: Audio-to-Audio \Sch{} Bridges}}.

The first challenge is curating a dataset that is both expansive enough to cover most genres of music of interest and being permissively licensed. To achieve this, we collected and filtered permissively licensed music data from public datasets, leading to 2.3K hours in total. As data quality varies significantly across datasets, we adopt the common pre-training and fine-tuning approach \citep{ouyang2022training}.

The second challenge is to support both restoration tasks in a \textit{single} model. We frame both tasks as the generative spectrogram inpainting task: bandwidth extension as inpainting the high-frequency part of the spectrogram along the frequency axis, and audio inpainting as frame inpainting along the time axis. The rest of the spectrogram should exactly match the input. We adopt the \Sch{} Bridge formulation \citep{liuI2SB,albergo2023stochastic} as it is particularly suitable for our generative spectrogram inpainting setup.

The third and most significant challenge is to train an \textit{end-to-end} model without using a vocoder or a codec. While prior works such as \citep{richter2023sgmse, jukic2024schr, ku2024gen} found success training directly on the complex spectrogram for speech enhancement, we find this ineffective when we need to synthesize missing spectrogram data. We use a factorized audio representation with power compression of the magnitude and trigonometric representation of the phase. We additionally apply phase orthogonalization based on the solution of the Procrustes problem to ensure that the generated phase values are consistent. These techniques make \modelname the first end-to-end high-res music restoration model. It is also advantageous over prior works that restore only the magnitude spectrum and apply a vocoder \citep{liu2024audiosr,liu2023maid}, as we preserve the original phase values where available.

The fourth challenge is to apply our restoration model -- which is trained on a fixed segment length -- to very long audio inputs. Directly concatenating individually restored segments could produce boundary artifacts especially for the bandwidth extension task. To solve this problem, we adapt MultiDiffusion \citep{bar2023multidiffusion}, which was originally designed for image panorama generation, to our model. We apply MultiDiffusion in overlapping sliding windows along the audio time axis, which allows \modelname to produce coherently restored outputs for hour-long sequences.

We demonstrate the effectiveness of our \modelname on several out-of-distribution test sets. Our model outperforms state-of-the-art baselines on these benchmarks. We also demonstrate the effectiveness of our factorized audio representation, phase orthogonalization, and inference methods qualitatively and quantitatively.
We summarize our contributions as follows:

\begin{enumerate}
    \item We propose \modelname, a state-of-the-art, end-to-end (vocoder-free), and multi-task diffusion \Sch{} Bridge model for 44.1kHz high-res music restoration.
    \item We propose an effective factorized audio representation and a phase orthogonalization method to achieve end-to-end training and generation. 
     \item We curate a large collection of permissively licensed public datasets for high-res music restoration, and design a training strategy for better utilization of the data.
    \item \modelname can coherently restore hour-long audio without boundary artifacts.
\end{enumerate}

\begin{figure}[!t]
    \centering
    \begin{tikzpicture}
        
        \node (A) {\includegraphics[width=3cm]{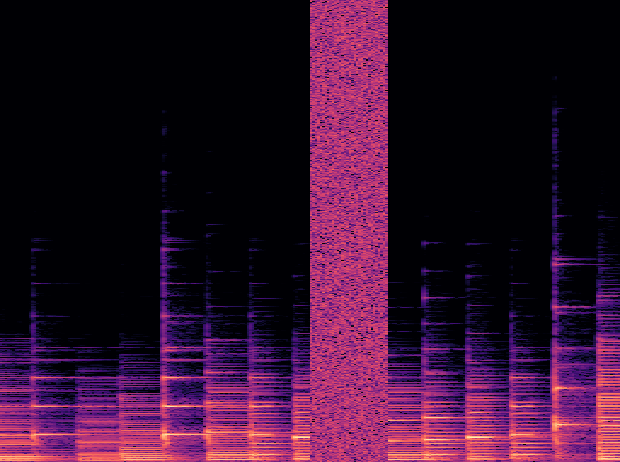}};
        \node (B) [right=0.2cm of A]
        {\includegraphics[width=3cm]{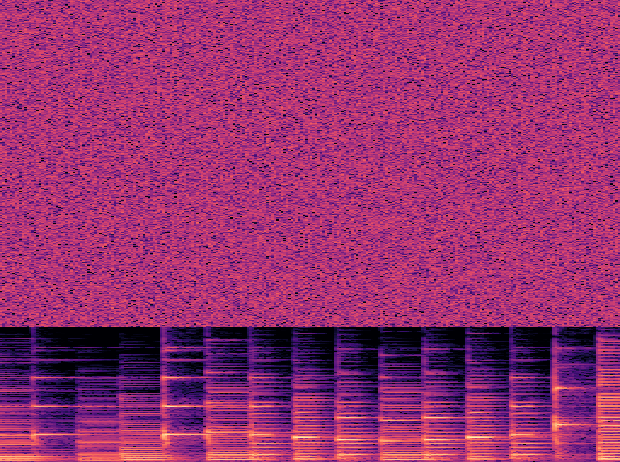}};
        \node (C) [right=2.5cm of B]
        {\includegraphics[width=3cm]{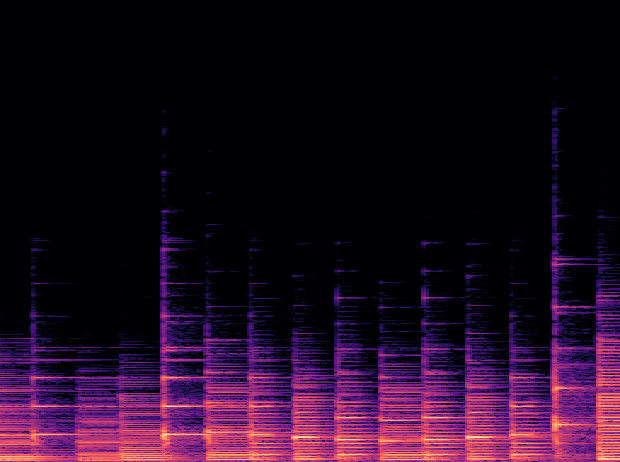}};
        
        \node at (A.south) [yshift=-0.2cm] {Inpainting};
        \node at (B.south) [yshift=-0.2cm] {Bandwidth Extension};
        \node at (C.south) [yshift=-0.2cm] {Clean};
        
        \node[left=0.01cm of A, align=center] (LeftBrace) 
            {\huge$\left\{\rule{0cm}{1.5cm}\right.$};
        \node[right=0.01cm of B, align=center] (RightBrace) 
        {\huge$\left.\rule{0cm}{1.5cm}\right\}$};

        \node (M) at ($(B)!0.56!(C)$) [xshift=-0.1cm, yshift=0.3cm] {\modelname};
    
        \draw [->, >=Stealth, ultra thick] (RightBrace.east) to (C.west);
    \end{tikzpicture}
    \caption{\modelname targets music restoration with a focus on inpainting and bandwidth extension, each corresponding to a specific corruption pattern in spectrogram. The model is then trained to fit the diffusion \Sch{} Bridge process from the corrupted distribution to the clean distribution.}
    \label{fig:simpleoverview}
\end{figure}

\section{Related Works}

\begin{table}[!t]
\centering
\small
\caption{Comparison between our \modelname with prior works on music restoration models. In addition to having better generation quality, our method is end-to-end without a vocoder or a codec, supports multiple restoration tasks, applies to high sampling rate, and supports restoration for very long audio. We aim to provide a detailed study of our proposed approaches for achieving these properties, and note that many of our solutions can be incorporated into prior frameworks. $\spadesuit$ \citet{liu2024audiosr} and \citet{liu2023maid}; $\blacklozenge$ \citet{wang2023audit}; $\clubsuit$ \citet{moliner2023solving} and \citet{moliner2023diffusion}.}\label{table: models}
\begin{tabular}{lcccc}
\toprule

Model & End-to-end & Multiple tasks & Sampling rate & Long audio restoration \\ \midrule

Conditional diffusion model $^\spadesuit$ & \xmark & \xmark & 48kHz & \xmark \\
Instruction based model $^\blacklozenge$ & \xmark & \cmark & 22.05kHz & \xmark \\
Inverse method $^\clubsuit$ & \cmark & \cmark & 22.05kHz & \xmark \\ \midrule
\rowcolor{pink!20}
\modelname & \cmark & \cmark & 44.1kHz & \cmark \\ \bottomrule
\end{tabular}
\end{table}

\subsection{Diffusion Models}

Diffusion models and score-based generative models \citep{sohl2015deep,ho2020denoising,songscore} are a type of generative models that aim to learn the data distribution $\pdata$ from data $\mathcal{X}$ via a forward and a backward stochastic differential equation (SDE). The forward SDE is
\begin{equation}
    dX_t=f_t(X_t)dt+\sqrt{\beta_t}dW_t,
\end{equation}
where $t\in[0,1]$ is the diffusion time index such that $X_0\sim\pdata$ and $X_1\sim\mathcal{N}(0,I)$, $f_t$ is an optional drift term linear in $X_t$, and $W_t$ is the Wiener process \citep{anderson1982reverse}. The reverse SDE can be analytically solved as
\begin{equation}
    dX_t = [f_t(X_t) - \beta_t \nabla_{X_t} \log p(X_t, t)]dt+\sqrt{\beta_t}d\bar{W}_t,
\end{equation}
where $p(X_t, t)$ is the marginal distribution of the forward SDE at time $t$, $\nabla_{X_t} \log p$ is its score function, and $\bar{W}_t$ is the reverse Wiener process. Then, a neural network $\epsilon(X_t,t)$ is trained to predict the score function, making it feasible to numerically sample from the approximated data distribution.

Diffusion models can be easily extended to conditional generation by conditioning $X_t$ on the condition $y$ everywhere in the forward and reverse SDEs. In practice, this is done by adding $y$ as an input to the neural network $\epsilon$. Different conditioning mechanisms have been proposed, such as concatenation at the input \citep{saharia2022image} or cross attention at each block \citep{rombach2022high}.
While this is a flexible framework for conditional generation, the generation quality heavily depends on the empirical conditioning mechanism, and there is no explicit regularization for the output to be faithful to the condition. This casts challenges to controllable generalization especially in data restoration.

\subsection{Inverse Problems}

Let $\mathcal{A}$ be a degradation mechanism (e.g., temporal masking or bandwidth limitation) and $y=\mathcal{A}(X_0)$ be the degraded observation of the clean data $X_0$. Inverse problems aim to restore $X_0$ from $y$ without training a conditional model. When $\mathcal{A}$ is affine, this becomes the linear inverse problem, and many works have proposed to modify the inference algorithm using a pre-trained unconditional model
\citep{ho2022video,chung2022diffusion,chung2022improving,song2023pseudoinverse}. In detail, one could estimate the score $\nabla_{X_t} \log p(y|X_t,t)$ and add it to the original score $\nabla_{X_t} \log p(X_t,t)$ similar to classifier guidance:
\begin{equation}
    \nabla_{X_t} \log p(X_t,t|y) = \nabla_{X_t} \log p(X_t,t) + \nabla_{X_t} \log p(y|X_t,t).
\end{equation}
The major advantage of this approach is that the same unconditional model could be applied to many restoration tasks as long as $\mathcal{A}$ is known and affine. The output is also regularized to be faithful to $y$. However, the generation quality heavily depends on the quality of the unconditional model (which is more challenging to train than conditional generation) and the estimation quality of $\nabla_{X_t} \log p(y|X_t,t)$.

\subsection{\Sch{} Bridges}

\Sch{} Bridges \citep{de2021diffusion,chen2021likelihood,liuI2SB,albergo2023stochastic} are a different approach to solve inverse problems. These models generalize diffusion models to the setting where $X_1$ is drawn from a pre-defined $p_{\text{deg}}(X_1|X_0)$ instead of the standard normal distribution. For instance, $p_{\text{deg}}$ can be a Gaussian distribution with mean $\mathcal{A}(X_0)$. The forward SDE is
\begin{equation}
    dX_t=[f_t(X_t)+\beta_t\nabla_{X_t}\log\Psi(X_t,t)]dt+\sqrt{\beta_t}dW_t,
\end{equation}
and the backward SDE is
\begin{equation}
    dX_t=[f_t(X_t)-\beta_t\nabla_{X_t}\log\hat{\Psi}(X_t,t)]dt+\sqrt{\beta_t}d\bar{W}_t,
\end{equation}
where $\Psi,\hat{\Psi}$ are energy potentials that solve certain coupled PDEs \citep{schrodinger1932theorie,leonard2013survey}. In \citet{liuI2SB}, the drift term is defined as $f_t=0$, leading to an analytic solution to the posterior distribution $q(X_t | X_0, X_1) = \mathcal{N}(X_t;\mu_t, \Sigma_t)$ under the Dirac delta assumption (see \eqref{eq: SB posterior}). Then, a neural network $\epsilon(X_t,t)$ is trained to predict $(X_t-X_0)$ up to a scaling factor. Sampling from the reverse SDE is similar to the diffusion models, except that one starts from a given $X_1$ instead of the random noise. \Sch{} Bridges embody the advantages of both conditional diffusion models and inverse algorithms: (1) the generation quality is similar to conditional diffusion model as it is trained on paired data, and (2) the generation is faithful to $X_1$ as the reverse SDE starts with $X_1$. These make \Sch{} Bridges suitable for data restoration, especially when $X_1$ is in part aligned with $X_0$. Therefore, we leverage the \Sch{} Bridge framework for audio restoration.

\subsection{Audio Restoration}

Audio restoration problems are inverse problems in the audio domain, where the degradation function is audio-specific including but not limited to low-pass filtering (or down-sampling), masking temporal segments, adding noise, and so on (see Table 3 in \citet{serra2022universal} for a more comprehensive summary of degradations). The corresponding tasks for these corruptions are called bandwidth extension (up-sampling), inpainting, denoising, etc. Our work focuses on music bandwidth extension and inpainting at 44.1kHz, as these are more challenging tasks than speech restoration tasks, which are often tackled at 16kHz.

Many diffusion-based approaches have been proposed for audio restoration. Most existing works are designed for speech restoration, where they use conditional diffusion models to restore degraded magnitude, and then apply a vocoder to produce the waveform \citep{serra2022universal,yu2023conditioning,moliner2023diffusion,kim2024audio,yun2025flowhigh}. There are a few models for music restoration: NuWave \citep{lee2021nu} and AudioSR \citep{liu2024audiosr} for bandwidth extension, and MAID \citep{liu2023maid} for inpainting. Moreover, with the recent success of multimodal models, instruction-based models have been proposed for audio restoration \citep{wang2023audit}, where the specific restoration task can be defined by text instructions.

Several other works are based on the inverse problem framework. CQT-Diff \citep{moliner2023solving} is a linear inverse problem approach for audio restoration. It trains an end-to-end 22.05kHz unconditional model with constant-Q transform (CQT) instead of short-term Fourier transform (STFT), and then applies the inverse algorithms proposed by \citet{ho2022video} for restoration. \citet{moliner2023diffusion} improved their architecture and applied to inpainting with a gap of up to 300 ms. We refer to \citet{lemercier2024diffusion} for a comprehensive survey of diffusion models for audio restoration.
There are also several orthogonal works using \Sch{} Bridges for speech enhancement \citep{jukic2024schr,wang2024diffusion, li2025bridge}.

Our work targets at the challenging music bandwidth extension and inpainting tasks at 44.1kHz. We use \Sch{} Bridges and achieve significantly better restoration quality. Our model is end-to-end requiring no vocoder, supports these restoration tasks with a single model, applies to high sampling rate, and can restore very long audio.
Table \ref{table: models} summarizes the comparison between our approach and existing works in the music restoration domain.

\section{Method}

Our \modelname is an end-to-end approach for music restoration at 44.1kHz that does not require a pre-trained vocoder or audio codec. We achieve this by training on an effective factorized audio representation (see Section \ref{sec: method: magphase}). We then train a \Sch{} Bridge model for music restoration based on \citet{liuI2SB}, with specific alterations for handling our audio representation (see Section \ref{sec: method: SB model}). Since phase predictions are free-form and may not be proper, Section \ref{sec: method: inversion} describes how we obtain waveform with the model outputs. Section \ref{sec: method: architecture} describes our network architecture. Section \ref{sec: method: training} describes our two-stage training paradigm. Section \ref{sec: method: sampling} describes an inference algorithm that supports arbitrarily long audio inputs without boundary artifacts.

For notations, we let $\tilde{X}\in[-1,1]^L$ be the 1-D raw waveform of clean audio with length $L$, and $X_t$ be the audio representation that we will use in the \Sch{} Bridge model at time $t$ with respect to the stochastic process. We ignore the subscript $t$ when there is no ambiguity.

\subsection{Magnitude-Phase Factorized Audio Representation}
\label{sec: method: magphase}

The short-time Fourier transformation (STFT) representation of $\tilde{X}$, $S=\mathrm{STFT}(\tilde{X})$, is a complex matrix in $\mathbb{C}^{N \times W}$, where $N$ is the number of frequency subbands and $W$ is the number of overlapping STFT frames. \footnote{We assume a 44.1kHz sampling rate, with hop size = 512, window length = 2048, and FFT
bins = 2048. We train with $W=256$, which corresponds to about 2.97 seconds of audio.} For simplicity, we can represent the complex values with their real and imaginary parts $[\mathrm{Re}(S), \mathrm{Im}(S)]$, leading to the two-channel spectrogram $S \in \mathbb{R}^{N \times W \times 2}$.
While existing vocoder-free methods directly model this two-channel representation \citep{richter2023sgmse,jukic2024schr,scoredec}, we factorize $S$ into magnitude and phase components of the STFT in our method. We find that separating them is necessary for the following reasons: 
\begin{enumerate}
    \item Magnitudes in adjacent frequency bands are strongly correlated, but this is less true for phase. Keeping them entangled would make it harder for the model to capture smooth changes in magnitude across frequencies. 
    \item The periodicity of phase makes fitting to it a more challenging task. In addition, phase estimation involves computing the ratio of the real and imaginary components, both of which would be near-zero floating point values at low magnitudes. 
    \item Phase-magnitude factorization limits the extent to which complications from fitting to the phase affect the magnitude estimation.
\end{enumerate}

To factorize $S$ into magnitude $\Lambda$ and phase $\Theta$ we compute:
\begin{equation}
\left\{
\begin{array}{rl}
    \Lambda_{i,j} & \displaystyle \coloneqq \sqrt{(S_{i,j,1}) ^2 + (S_{i,j,2}) ^2} \\
    \Theta_{i,j} & \displaystyle \coloneqq \mathrm{atan2}(S_{i,j,2}, S_{i,j,1})
\end{array}
\right..
\end{equation}

We further use the trigonometric representation to model the phase $\Theta$ \citep{peer2022phase}. The trigonometric projections can be interpreted as the original two-channel spectrogram $S$ with normalized magnitude (i.e., all magnitudes $=1$).
Our final factorized representation $X\in \mathbb{R}^{N\times W \times 3}$ is defined as:
\begin{equation}
\label{eq: audio representation}
\left\{
\begin{array}{rl}
    X_{i,j,1} & \displaystyle \coloneqq (\Lambda_{i,j}) ^ {\rho}\\
    X_{i,j,2} & \displaystyle \coloneqq \cos(\Theta_{i,j}) \\
    X_{i,j,3} & \displaystyle \coloneqq \sin(\Theta_{i,j})
\end{array}
\right..
\end{equation}
We set $\rho < 1$ to compress the range of the magnitude values, using $\rho\coloneqq 0.25$ in our experiments. Our experiments in \ref{sec: experiments: factorization} find that the factorized representation results in a better fit to the data distribution than 
the un-factorized $S$.

\subsection{Music Restoration with \Sch{} Bridges}
\label{sec: method: SB model}

We train a \Sch{} Bridge model on the three-channel representation $X$ described in \eqref{eq: audio representation}. Following \citet{liuI2SB}, we let $X_0\in\mathbb{R}^{N\times W\times 3}$ be the clean sample inputs, and $X_1$ be degraded samples. We focus on bandwidth extension and inpainting, both of which can be formulated as the masking corruption similar to image inpainting. Let $\mathbb{M}\in\mathbb{B}^{N \times W \times 3}$ be the boolean mask for masking, where $\mathbb{B}=\{0,1\}$. For bandwidth extension, $\mathbb{M}_{i,j,k}=1$ for $i> N'$, where $N'$ refers to the highest subband in the degraded audio. 
$N'$ is randomly sampled from subbands representing frequencies above 4kHz. For inpainting, $\mathbb{M}_{i,j,k}=1$ for $W_1\leq j\leq W_2$, where $W_1$ and $W_2$ refer to the starting and ending frame of missing audio. Following \citet{liu2023maid}, we randomly sample $W_1$ and $W_2$ such that the inpainting gap is uniform between 0.1 and 1.6 seconds. For a certain mask $\mathbb{M}$, we define $X_1$ as
\begin{equation}
\label{eq: degraded}
    X_1 = X_0 \odot (\mathbbm{1}-\mathbb{M}) + \eta_{\text{fill}}  \odot \mathbb{M},
\end{equation}
where $\odot$ refers to the element-wise product, and $\eta_{\text{fill}} \sim \mathcal{N}(0, \sigma_{\text{fill}}^2 I)$ in order to define a Gaussian $p_{\text{deg}}(X_1|X_0)$ for the masked area in our audio representation. Note that if $\mathbb{M}=\mathbbm{1}$ and $\sigma_{\text{fill}}=1$, the \Sch{} Bridge degenerates to an unconditional diffusion model, where $X_1$ is Standard Normal.

According to \citet{liuI2SB}, $X_t$ is sampled from $q(X_t | X_0, X_1) = \mathcal{N}(X_t;\mu_t, \Sigma_t)$, where $\mu_t$ and $\Sigma_t$ equal to
\begin{equation}
\label{eq: SB posterior}
    \mu_t = \frac{\bar{\sigma}_t^2X_0+\sigma_t^2X_1}{\bar{\sigma}_t^2+\sigma_t^2},
    \Sigma_t = \frac{\bar{\sigma}_t^2\sigma_t^2I}{\bar{\sigma}_t^2+\sigma_t^2},
\end{equation}
where $\sigma_t^2=\int_0^t\beta_\tau d\tau,\bar{\sigma}_t^2=\int_t^1\beta_\tau d\tau$. The $\beta_t$ schedule is symmetric on each side of $t=\frac12$: $\beta_t=\beta_{1-t}=\min(t,1-t)^2\cdot\beta_{\max}$, where $\beta_{\max}$ is a hyper-parameter to tune. In our model, we use $\beta_{\max}=1$ as it leads to the most stable training. The training objective is only computed on masked region of our audio representation:

\begin{equation}
\label{eq: training objective}
\mathcal{L}(\epsilon) = \mathbb{E}_{t\sim \mathcal{U}([0,1]), \mathbb{M}, X_0 \sim \mathcal{X}, X_t \sim q(X_t| X_0, X_1)}\left \|\mathbb{M}\odot\left(\epsilon(X_t, t) - \frac{X_t - X_0}{\sigma_t}\right) \right \|_2^2,
\end{equation}

where 
$t$ is sampled from the uniform distribution on $[0,1]$, 
$X_0$ is sampled from $\mathcal{X}$, 
$X_t$ is sampled from $q(X_t| X_0, X_1)$,
$\epsilon$ is the neural network taking both $X_t$ and step $t$ as inputs, 
and $\mathbb{M}$ is uniformly sampled from the bandwidth extension and inpainting masks.

\subsection{Waveform Synthesis with Phase Orthogonalization}
\label{sec: method: inversion}

All operations defined in Section \ref{sec: method: magphase} are invertible and should allow us to recover the original 1-D waveform signal almost exactly. We can reconstruct the two-channel spectrogram $\hat{S}$ from $X$ with:
\begin{equation}
\left\{
    \begin{array}{rl}
    \hat{S}_{i,j,1} &= X_{i,j,2} \cdot (X_{i,j,1})^{1/\rho} \\
    \hat{S}_{i,j,2} &= X_{i,j,3} \cdot (X_{i,j,1})^{1/\rho}
    \end{array}
\right..
\end{equation}
Then, applying the inverse STFT with the same STFT parameters yields the waveform.

However, when sampling from our trained neural network, we cannot guarantee that the unconstrained model outputs $[X_{i,j,2}, X_{i,j,3}]$ satisfy the trigonometric representation of phase: $X_{i,j,2}^2+ X_{i,j,3}^2=1$. This could manifest as an additional scaling of the reconstructed spectrogram $S$, which is undesirable. To alleviate this issue, we use phase orthogonalization to map $[X_{i,j,2}, X_{i,j,3}]$ to the last-squares-nearest valid configuration. This is in part inspired by the analysis in \citet{levinsonSVD} for learning 3D rotations, though we require only the 2D rotations in our case. Furthermore, the least-squares optimality of SVD orthogonalization is ideal for the removal of small amounts of Gaussian noise, making it compatible with the Gaussian diffusion process. Approaches such as \citet{chen2023riemannian} can also guarantee proper rotation values, but we find our approach to be simple and practical enough for our use case.

Let $\hat{R}_{i,j} \in \mathbb{R}^{2 \times 2}$ be a noisy estimate of a rotation matrix at spectrogram coordinate $(i,j)$, which is constructed with
\begin{equation}
    \hat{R}_{i,j} \coloneqq \begin{bmatrix}
X_{i,j,2} & -X_{i,j,3} \\
X_{i,j,3} &  X_{i,j,2}
\end{bmatrix}.
\label{eq: rotation-matrix}
\end{equation}

We then compute its nearest valid configuration in least squares as follows:
\begin{equation}
\label{eq: orthogonalization}
    \texttt{SVDO}^+(\hat{R}_{i,j}) \coloneqq \argmin_{R_{i,j} \in \text{SO}(2)} \| R_{i,j} - \hat{R}_{i,j} \|_F ^2,
\end{equation}
where $\text{SO}(2)$ is the orthogonal group in two dimensions. Note that for any $2 \times 2$ matrix $A$, we have the following solution \citep{levinsonSVD,schonemann1966generalized}:
\begin{equation}
     \texttt{SVDO}^+(A) = U \Sigma' V, \text{ where } \Sigma' = \mathrm{diag}(1, \mathrm{det}(UV^{\top})),
\end{equation}
where $A=U\Sigma V$ is the SVD decomposition. Applying SVD to $\hat{R}_{i,j}$ yields
\begin{equation}
    U=(X_{i,j,2}^2+X_{i,j,3}^2)^{-\frac12}\hat{R}_{i,j}, \Sigma=(X_{i,j,2}^2+X_{i,j,3}^2)^{\frac12} I, V=I.
\end{equation}
And therefore, the solution is
\begin{equation}
    \texttt{SVDO}^+(\hat{R}_{i,j}) = (X_{i,j,2}^2+X_{i,j,3}^2)^{-\frac12}\hat{R}_{i,j}
\end{equation}
as $\mathrm{det}(UV^{\top})=1$. Then, the orthogonalized phase estimation allows us to reconstruct the spectrogram with
\begin{equation}
    \hat{S}_{i,j} = (X_{i,j,1})^{1/\rho} \cdot (\texttt{SVDO}^+(\hat{R}_{i,j}))_{:,1}.
\end{equation}

We further compute the minimum residual as
\begin{equation}
\label{eq: orthogonalization error}
    \mathbf{Err}_{\text{phase-ortho}}(X_{i,j}) = \| \texttt{SVDO}^+(\hat{R}_{i,j}) - \hat{R}_{i,j} \|_F ^2 = 2\left((X_{i,j,2}^2+X_{i,j,3}^2)^{\frac12}-1\right)^2.
\end{equation}
We will quantify and visualize this error in our experiments. We expect a well trained model should have a small error, and the phase orthogonalization only does minor correction. \footnote{We additionally note that manifold generative models such as \citet{chen2023riemannian} could also address this issue without orthogonalization, yet we find our approach simple and effective enough and therefore leave this approach for future work.}

\subsection{Architecture}
\label{sec: method: architecture}
Our model closely follows the conditional UNet architecture as commonly used in prior works \citep{ronneberger2015u,dhariwal2021diffusion,liuI2SB}, with some modifications. Notably, absolute positional embedding layers were replaced with 2-D rotary position embedding (RoPE) \citep{su2024roformer}. Further, we use an additional conditioning variable $C\in\mathbb{R}^{N\times W}$ via absolute positional embeddings. $C$ only varies in the frequency axis: $C_{i,j}=i$, $1\leq i\leq N$. This allows the model to strongly condition on the frequency, while maintaining translational equivariance along the temporal axis in the spectrogram.

In terms of the neural network configuration, there are five up-sampling and down-sampling layers, each having two residual blocks. The hidden channels are [128, 256, 512, 768, 1024, 2048]. Both input and output have three channels, following our audio representation in~\eqref{eq: audio representation}. The diffusion step embedding dimension is 128, following \citet{kongdiffwave}. The network has 565M parameters.

\subsection{Two-Stage Training}
\label{sec: method: training}

We follow the common pre-training and fine-tuning approach for stable large scale training \citep{ouyang2022training}. During pre-training, we train our \Sch{} Bridge model from scratch on 2.3K hours of training data. We use \texttt{bf16} for more efficient training. During fine-tuning, we train on a 1.5K-hour high quality subset and use \texttt{fp32}, ensuring the model produces clean and meaningful sound for corrupted parts.

During pre-training, we use as much data as possible to train our model from scratch. We use \texttt{bfloat16} for more efficient training. During fine-tuning, we only train on high quality subsets of the data with \texttt{fp32}, ensuring the model produces clean and meaningful sound for corrupted parts.
In detail, the fine-tuning set includes:
FMA \citep{defferrard2016fma},
Medley-solos-DB \citep{lostanlen2016deep},
MTG-Jamendo \citep{bogdanov2019mtg},
Musan \citep{snyder2015musan},
Music Instrument \citep{musicinstrument},
MusicNet \citep{thickstun2017learning}, and
Slakh \citep{manilow2019cutting}.
The pre-training set additionally includes:
CLAP-Freesound \citep{wu2023large},
GTZAN \citep{sturm2013gtzan},
MusicCaps \citep{agostinelli2023musiclm},
NSynth \citep{engel2017neural},
PianoTriads \citep{pianotriads}.
We carefully examined all data licenses in these datasets and only selected the permissively licensed audio to train our model (i.e., we removed data that are NC, ND, SA, or under unknown licenses, etc.).

During fine-tuning, we adopt the $t$-range partitioning strategy from \citet{balaji2022ediff}: we fine-tune separate models on different $t$ intervals, each initialized from the same pre-trained checkpoint. This leads to models specialized in different noise level ranges. We choose the intervals that partition noise level ranges between $\sigma_0^2$ and $\sigma_1^2$. In 2-partitioning, the intervals are $t\in(0,\frac12]$ and $t\in[\frac12,1]$; in 4-partitioning, the intervals are $t\in(0,\frac{1}{2^{4/3}}]$, $t\in[\frac{1}{2^{4/3}},\frac12]$, $t\in[\frac12,1-\frac{1}{2^{4/3}}]$, and $t\in[1-\frac{1}{2^{4/3}},1]$.
During sampling, we use the corresponding checkpoint based on the exact $t$.

\subsection{Sampling}
\label{sec: method: sampling}

\begin{algorithm}[!t]
\caption{MultiDiffusion sampling at step $t$}
\label{alg:mdiff}

\KwIn{$\mathbf{X}_t^{\mathrm{full}}\in\mathbb{R}^{N\times W^{\mathrm{full}} \times 3}$, $t$, $W$, $H$, $\epsilon(\cdot,\cdot)$}
\tcc{Input noisy spectrogram tensor, sampling timestep, window size, MultiDiffusion hop size, model. Tensors are in bold.}

$\mathbf{C}, \mathbf{V} \gets \mathbf{0} \in \mathbb{R}^{N\times W^{\mathrm{full}} \times 3} $\;
\tcc{Initialize normalizing \& output tensor.}

$j \gets 0$\;
\tcc{window left-most position index (numpy-style indexing)}
\While{$j + W < W^{\mathrm{full}}$}{
$\mathbf{X}_t^{\mathrm{patch}} \gets \mathbf{X}_t^{\mathrm{full}}[:,j:(j+W),:]$\;
\tcc{Create a patch.}

\tcc{Accumulate model output and normalizing tensors.}

$\mathbf{V}[:,j:(j+W),:] \gets \mathbf{V}[:,j:(j+W),:] + \epsilon(\mathbf{X}_t^{\mathrm{patch}},t)$\;

$\mathbf{C}[:,j:(j+W),:] \gets \mathbf{C}[:,j:(j+W),:]+\mathbf{1}$\;

$j \gets j + H$\;
\tcc{Shift processing window by hop size.}
}

\Return{$\mathbf{V} \oslash \mathbf{C}$}\;
\tcc{Average with element-wise division.}

\end{algorithm}

The sampling algorithm given $X_1$ directly follows the diffusion model \citep{ho2020denoising}. Let $\Delta t$ be a step size where $\frac{1}{\Delta t}$ is an integer referring to the number of sampling steps. There is an analytic form for the posterior (see proof of Proposition 3.3 in \citet{liuI2SB}):
\begin{equation}
\label{eq: sampling}
      p(X_{t-\Delta t} | X_0, X_{t}) = \mathcal{N}\left(\frac{(\Delta\sigma_t^2)X_0 + \sigma_t^2 X_t}{\Delta\sigma_t^2 + \sigma_t^2}, \frac{(\Delta\sigma_t^2)  \sigma_t^2}{\Delta\sigma_t^2 + \sigma_t^2}I\right),
\end{equation}
where $\Delta\sigma_t^2\coloneqq\sigma_t^2 - \sigma_{t-\Delta t}^2$. During sampling, the $X_0$ is replaced with the predicted $X_0\coloneqq X_t - \sigma_t\epsilon(X_t,t)$ as indicated by \eqref{eq: training objective}. Then, repeating \eqref{eq: sampling} for $\frac{1}{\Delta t}$ steps yields the final output.

In practice, the audio we would like to up-sample or inpaint may be much longer than our training segment length. This is similar to the panorama generation problem in image generation, which could be solved by MultiDiffusion \citep{bar2023multidiffusion}. Inspired by their approach, we apply MultiDiffusion to extend our sampling process to arbitrary length.
Our algorithm is similar to Algorithm 2 in \citet{bar2023multidiffusion}, where the condition is the degraded audio in our case.

Formally, let $X_t^{\mathrm{full}}\in\mathbb{R}^{N\times W^\mathrm{full}\times3}$ be a degraded sample of arbitrary length that we would like to up-sample or inpaint. For simplicity, we assume $W^\mathrm{full}=K W$ where $K>1$. \footnote{The arbitrary length case can be easily tackled with padding in the end.} Our trained model $\epsilon(\cdot, t$) can process inputs of size $N \times W \times 3$ where $W$ corresponds to $256$ STFT frames (2.97 seconds). At diffusion time $t$, we compute the model's output on the full sample $\epsilon(X_t^{\mathrm{full}},t)$ as follows. We process our input $X_t^{\mathrm{full}}$ with a sliding window of width $W$ and shifting the position by a hop size $H < W$ (typically $128$ for $50\%$ overlap) until all of $X_t$ is processed. Outputs in overlapping areas are uniformly averaged, though other weighting functions are topic of future work \citep{polyak2024movie}. Cyclic padding is used to ensure the last input window has a full temporal width of $W$.
Our MultiDiffusion processing is detailed in \eqref{eq: multidiffusion} and the pseudo code is in Algorithm \ref{alg:mdiff}.

\begin{equation}
\label{eq: multidiffusion}
    \begin{array}{rl}
        \mathbf{int}_k
        & \displaystyle = \left[\frac{k-1}{2}W:\frac{k+1}{2}W\right], k=1,\cdots,2K-1; \\
        \mathbf{v}_k
        & \displaystyle = \epsilon\left(\mathbf{crop}(X_t^{\mathrm{full}}, \mathbf{int}_k), t\right), k=1,\cdots,2K-1; \\
        \epsilon(X_t^{\mathrm{full}},t)
        & \displaystyle \coloneqq \left(\sum_{k=1}^{2K-1}\mathbf{pad}(\mathbf{v}_k,\mathbf{int}_k)\right) \mathbin{\oslash} \left(\sum_{k=1}^{2K-1}\mathbf{pad}(\mathbf{1}_{N\times W\times3}, \mathbf{int}_k)\right),
    \end{array}
\end{equation}
where
\begin{itemize}
    \item $\mathbf{int}_k$ refers to the $k$-th interval in the sliding-window segmentation of the full sample with $W/2$ overlapping frames between consecutive segments,
    \item $\mathbf{crop}$ refers to the cropping operation applied to the full sample given the interval indicating start and end frames,
    \item $\mathbf{pad}$ refers to the zero-padding (to shape of full sample) operation applied to a segment given the interval indicating start and end frames in the full sample (note that $\mathbf{pad}$ is a right inverse function of $\mathbf{crop}$),
    \item and $\mathbin{\oslash}$ is the element-wise division symbol (inverse function of $\odot$).
\end{itemize}

\section{Experiments}

\subsection{Experimental Setup}

\paragraph{\modelname.}
We train different configurations of our model as described in Section \ref{sec: method: training}: no partitioning, 2-partitioning, and 4-partitioning. \footnote{For the 4-partitioning model, we apply an additional loss mask to compute loss only up to the maximum frequency of each training segment to make training more stable.} For all of our models, we pre-train with a batch size of 320 for 209K iterations, and fine-tune with a batch size of 64 for additional 250K iterations. We use gradient clip $=0.5$ and learning rate $8\times10^{-5}$. The training segment length is 130560, which equals about 2.97 seconds of audio. We sample the area removal mask in \eqref{eq: degraded} uniformly from the inpainting mask and the bandwidth extension mask. We use 32 NVIDIA A100 GPUs to train our models. At inference time, we use 50 sampling steps for bandwidth extension, and 200 sampling steps for inpainting.

\paragraph{Baselines.}

For each of the bandwidth extension and inpainting task, we consider three baselines: conditional diffusion method, inverse method, and instruction-based method.

\begin{itemize}
    \item \textbf{Conditional diffusion.} We use \textit{AudioSR} \citep{liu2024audiosr} for bandwidth extension. AudioSR could extend audio to 48kHz with training segment length 5.12 seconds. We consider \textit{MAID} \citep{liu2023maid} for inpainting. MAID could inpaint audio at 44.032kHz with training segment length 131072. We re-train MAID on our training dataset.

    \item \textbf{Inverse method.} We consider \textit{CQTDiff} \citep{moliner2023solving} for both tasks. Since the original CQTDiff is small and trained on 22.05kHz, we re-train a larger CQTDiff on 44.1kHz using our training dataset. We increase the depth from 6 to 8 and double the channels to $[64, 128, 128, 256, 256, 256, 256, 256, 256]$, leading to a $5.75\times$ larger model. It is the largest model we find to have stable training in our experiments.

    \item \textbf{Instruction-based method.} Since existing instruction-based models \citep{wang2023audit} are only 24kHz or less, we train our own 48kHz instruction-based  baseline model on our training dataset with the following design. We use the instruction templates for up-sampling and inpainting from Audit \citep{wang2023audit}, an optimized diffusion transformer architecture \citep{peebles2023scalable} similar to \citet{lee2024etta}, the byT5 embedding \citep{xue2022byt5}, the optimal-transport conditional flow matching loss function \citep{lipman2022flow,tong2023conditional}, and a 48kHz BigVGAN-v2 vocoder \citep{lee2023bigvgan}. For conciseness, we name this model \textit{IBAR}: Instruction-Based Audio Restoration.
\end{itemize}

\paragraph{Evaluation setup.}
We evaluate all models on several 44.1kHz out-of-distribution (OOD) test sets:
\begin{itemize}
    \item AAM \citep{ostermann2023aam}: a collection of synthetic music. We randomly select 93 test samples for evaluation. Each sample is between 2 and 3 minutes.
    \item CCMixter \footnote{\url{https://ccmixter.org/}}: a collection of remixed music. We use the same set as \citet{liutkus2014kernel}. Each sample is between 1 and 6 minutes.
    \item MTD \citep{zalkow2020mtd}: a collection of classical music pieces. We randomly select 200 test samples for evaluation. Each sample is between 10 seconds and 1 minute.
\end{itemize}


Our bandwidth extension evaluation follows \citet{liu2024audiosr} and evaluates three cutoff frequencies: 4kHz, 8kHz, and 12kHz. We resample the ground truth audio to twice the cutoff frequency and use it as the input to all models. 
For the inpainting evaluation, we mask a fixed-length (300ms, 500ms, 1000ms) segment every 5 seconds. We then run the model with its receptive field centered on each masked region to inpaint the missing content.

For \textbf{objective evaluation metrics}, we report the following.

\begin{itemize}
    \item Log-spectral distance (LSD) \citep{erell1990estimation}, a spectrogram distance metric computed as 
    \begin{equation}
        \mathrm{LSD} = \frac{1}{W}\sum_{j=1}^W \left[ \frac{1}{N}\sum_{i=1}^N \left( \log_{10}\frac{\Lambda_{i,j}^2}{\hat{\Lambda}_{i,j}^2} \right)^2 \right]^{\frac12},
    \end{equation}
    where $\Lambda$ is the ground truth magnitude and $\hat{\Lambda}$ the magnitude of the model's prediction.

    \item Scale-invariant spectrogram-to-noise ratio (SiSpec) \citep{liu2021voicefixer}, a signal-to-noise ratio metric computed as 
    \begin{equation}
        \mathrm{SiSpec} = 10\cdot\log_{10}\frac{\|\mathbf{n}(\Lambda)\|^2}{\|\mathbf{n}(\Lambda)-\hat{\Lambda}\|^2},
    \end{equation}
    where $\mathbf{n}(\Lambda) = \langle\hat{\Lambda},\Lambda\rangle\Lambda/\|\hat{\Lambda}\|^2$ is the scale invariant normalization of the ground truth magnitude.

    \item ViSQOL \citep{chinen2020visqol}, an objective perceptual quality for 48kHz audio, which measures similarity scores by comparing the spectro-temporal features and maps to the Mean Opinion Score (MOS) scale between 1 and 5. The ground truth has a score of $4.732$.

\end{itemize}

In additon to the above evaluation datasets, we train and evaluate all models on the Maestro dataset \citep{hawthorne2018enabling}, a dataset of classic piano music. In addition to LSD and SiSpec, we compute the $\mathrm{F}_1$ score of MIDI transcriptions using the \texttt{mir\_eval} package \citep{raffelmir_eval}.

We conduct \textbf{human evaluation} on the bandwidth extension (cutoff = 4kHz) and inpainting (gap = 1000ms) experiments due to the limitations of objective metrics. For each OOD test dataset, we randomly select fifty segments and ask human listeners to rate the output quality based on how close they sound compared to the ground truth and report Mean Opinion Scores (MOS).

\subsection{Bandwidth Extension Results}

The bandwidth extension results are shown in Tables \ref{tab: AAM BWE main} - \ref{tab: Maestro BWE main}. We find our \modelname achieves better SiSpec in most cases, indicating it has the best signal-to-noise ratio (SNR) and therefore the least noise up to a scale transformation. \modelname also achieves consistently better ViSQOL, indicating the samples have perceptually better quality. Our instruction based audio restoration baseline (IBAR) achieves better LSD in most cases, indicating it models the absolute value of the magnitude most accurately. Nevertheless, \modelname still achieves very good LSD compared to other baselines, and especially achieves similar LSD as IBAR on AAM, without using a vocoder.

Comparing different $t$-range partitioning approaches in \modelname, we find the 2-partitioned model (on $t\in (0,\frac12]$ and $t\in[\frac12,1]$) yields the better results overall. We conclude that two models trained on each splitted $t$-range are more expert than the model trained on $t\in(0,1]$ without partitioning. We also find that having more partitions helps modeling classic music (MTD) especially on the perceptual quality, and therefore we report the 4-partitioning results on Maestro.

\begin{figure}[!t]
    \centering
    \includegraphics[width=0.8\linewidth]{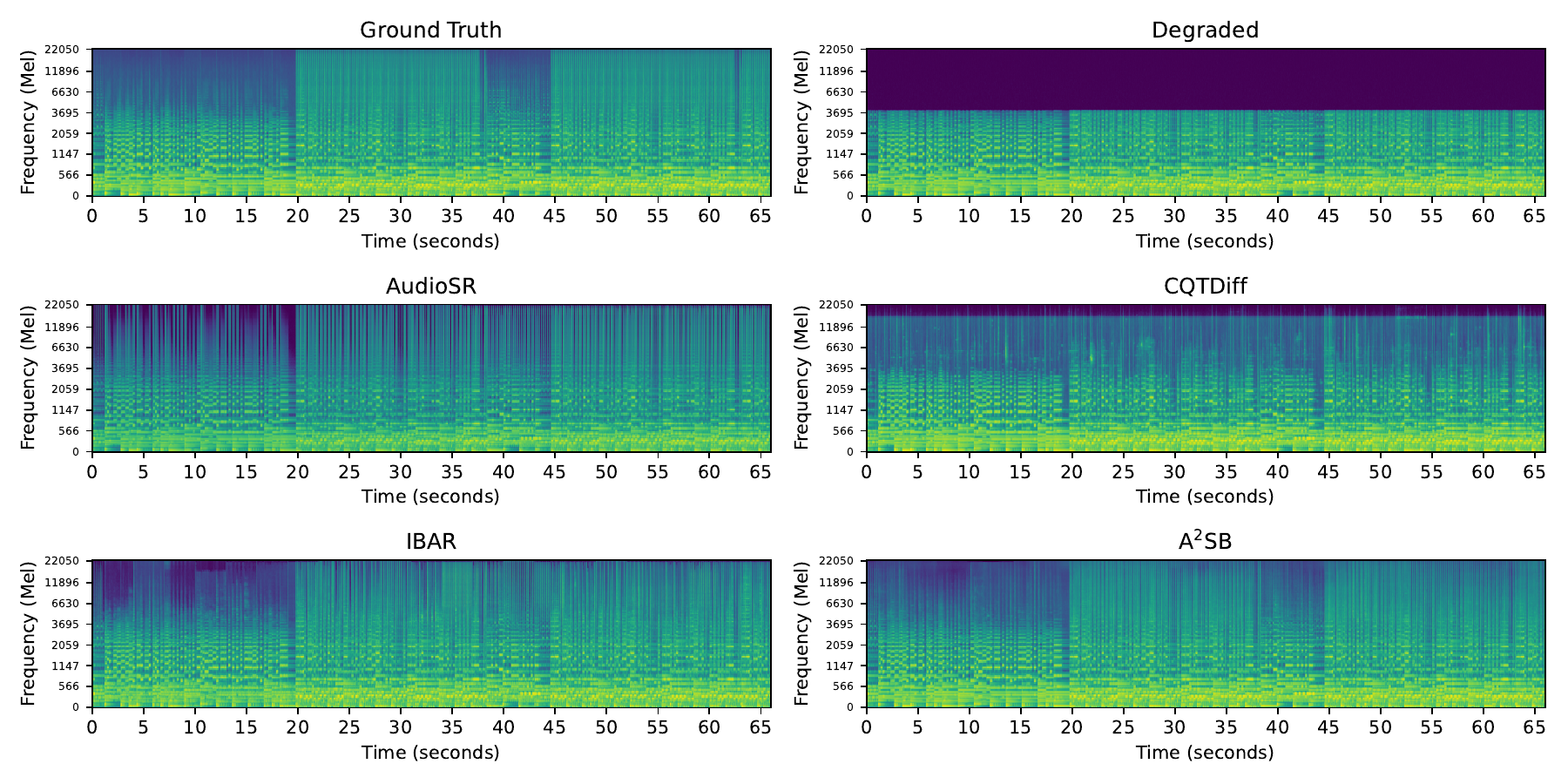}
    \caption{Qualitative comparison between different bandwidth extension methods with cutoff = 4kHz.}
    \label{fig: bwe example main}
\end{figure}

In Figure \ref{fig: bwe example main}, we show a qualitative sample of different bandwidth extension baselines. More samples can be found in Appendix \ref{app: BWE samples}. AudioSR often has artifacts around the cutoff frequency (Figures \ref{fig: app: bwe 5} and \ref{fig: app: bwe 6}), and it sometimes hallucinates on high frequency regions by adding percussion sounds (Figure \ref{fig: app: bwe 4}). CQTDiff usually has much worse quality. IBAR sometimes has unsmooth generations across frames, and it occasionally fails to produce a meaningful output. \modelname generates better quality overall, produces smoother and consistent content, and keeps the beats stable without hallucination on beats and percussion.

\begin{table}[!h]
    \centering
    \caption{Bandwidth extension results on AAM (synthetic music).}
    \tiny
    \begin{tabular}{l|S[round-precision=2] S[round-precision=2] S S[round-precision=2] S[round-precision=2] S S[round-precision=2] S[round-precision=2] S}
        \toprule
        \multirow{2}{*}{\textbf{Method}} 
        & \multicolumn{3}{c}{\textbf{Cutoff = 4kHz}} 
        & \multicolumn{3}{c}{\textbf{Cutoff = 8kHz}} 
        & \multicolumn{3}{c}{\textbf{Cutoff = 12kHz}} \\
        \cmidrule(lr){2-4} \cmidrule(lr){5-7} \cmidrule(lr){8-10}
         & {LSD $\downarrow$} & {SiSpec $\uparrow$} & {ViSQOL $\uparrow$} 
         & {LSD $\downarrow$} & {SiSpec $\uparrow$} & {ViSQOL $\uparrow$} 
         & {LSD $\downarrow$} & {SiSpec $\uparrow$} & {ViSQOL $\uparrow$} \\
        \midrule
        AudioSR
        & \num{2.222} & \num{13.725} & \num{3.057} & 
        \num{1.936} & \num{14.610}   & \num{3.455} & 
        \num{1.621} & \num{19.933}   & \num{3.783} \\
        CQTDiff
        & \num{2.367} & \num{19.905} & \num{1.926} & 
        \num{2.389} & \num{22.224}   & \num{1.928} & 
        \num{2.416} & \num{22.631}   & \num{1.965} \\
        IBAR
        & \bestnumber{\num{1.38}} & \num{8.51} & \num{2.951} & 
        \num{1.16} & \num{10.82}               & \num{3.384} & 
        \bestnumber{\num{0.991}} & \num{12.31} & \num{4.102} \\ \hline
        \rowcolor{pink!20} \modelname (no partitioning)
        & \num{1.40} & \num{19.28}             & \num{3.004} & 
        \bestnumber{\num{1.15}} & \num{27.35}  & \num{3.412} & 
        \bestnumber{\num{0.989}} & \num{31.33} & \num{3.947} \\
        \rowcolor{pink!20} \modelname (2-partitioning)
        & \num{1.44} & \bestnumber{\num{23.03}}            & \bestnumber{\num{3.248}} & 
        \bestnumber{\num{1.15}} & \bestnumber{\num{28.69}} & \num{3.706} & 
        \bestnumber{\num{0.99}} & \bestnumber{\num{31.76}} & \num{4.231} \\
        \rowcolor{pink!20} \modelname (4-partitioning)
        & \num{1.49} & \num{22.59} & \num{3.110} & 
        \num{1.20} & \num{28.46}   & \bestnumber{\num{3.773}} & 
        \num{1.04} & \num{31.67}   & \bestnumber{\num{4.340}} \\
        \bottomrule
    \end{tabular}
    \label{tab: AAM BWE main}
\end{table}

\begin{table}[!t]
    \centering
    \caption{Bandwidth extension results on CCMixter (remixed music).}
    \tiny
    \begin{tabular}{l|S[round-precision=2] S[round-precision=2] S S[round-precision=2] S[round-precision=2] S S[round-precision=2] S[round-precision=2] S}
        \toprule
        \multirow{2}{*}{\textbf{Method}} 
        & \multicolumn{3}{c}{\textbf{Cutoff = 4kHz}} 
        & \multicolumn{3}{c}{\textbf{Cutoff = 8kHz}} 
        & \multicolumn{3}{c}{\textbf{Cutoff = 12kHz}} \\
        \cmidrule(lr){2-4} \cmidrule(lr){5-7} \cmidrule(lr){8-10}
         & {LSD $\downarrow$} & {SiSpec $\uparrow$} & {ViSQOL $\uparrow$} 
         & {LSD $\downarrow$} & {SiSpec $\uparrow$} & {ViSQOL $\uparrow$} 
         & {LSD $\downarrow$} & {SiSpec $\uparrow$} & {ViSQOL $\uparrow$} \\
        \midrule
        AudioSR &
        \num{2.004} & \num{12.504} & \num{2.746} &
        \num{1.857} & \num{14.929} & \num{3.097} &
        \num{1.748} & \num{18.354} & \num{3.510} \\
        CQTDiff & 
        \num{2.011} & \num{14.674} & \num{1.970} &
        \num{2.060} & \num{15.875} & \num{1.860} &
        \num{2.097} & \num{16.337} & \num{1.850} \\
        IBAR & 
        \bestnumber{\num{1.64}} & \num{7.11}  & \num{2.373} & 
        \bestnumber{\num{1.41}} & \num{10.46} & \num{2.604} & 
        \bestnumber{\num{1.36}} & \num{7.86}  & \num{2.744} \\ \hline
        \rowcolor{pink!20} \modelname (no partitioning) & 
        \num{1.93} & \num{14.05} & \num{2.770} & 
        \num{1.71} & \num{19.95} & \num{3.200} & 
        \num{1.48} & \num{27.17} & \num{4.047} \\
        \rowcolor{pink!20} \modelname (2-partitioning) & 
        \num{1.85} & \bestnumber{\num{18.00}} & \bestnumber{\num{2.851}} & 
        \num{1.62} & \bestnumber{\num{23.39}} & \bestnumber{\num{3.438}} & 
        \num{1.45} & \bestnumber{\num{29.26}} & \num{4.211} \\
        \rowcolor{pink!20} \modelname (4-partitioning) & 
        \num{1.84} & \num{17.46} & \num{2.657} & 
        \num{1.65} & \num{23.17} & \num{3.430} & 
        \num{1.50} & \num{29.20} & \bestnumber{\num{4.234}} \\
        \bottomrule
    \end{tabular}
    \label{tab: CCMixter BWE main}
\end{table}

\begin{table}[!t]
    \centering
    \caption{Bandwidth extension results on MTD (classical music).}
    \tiny
    \begin{tabular}{l|S[round-precision=2] S[round-precision=2] S S[round-precision=2] S[round-precision=2] S S[round-precision=2] S[round-precision=2] S}
        \toprule
        \multirow{2}{*}{\textbf{Method}} 
        & \multicolumn{3}{c}{\textbf{Cutoff = 4kHz}} 
        & \multicolumn{3}{c}{\textbf{Cutoff = 8kHz}} 
        & \multicolumn{3}{c}{\textbf{Cutoff = 12kHz}} \\
        \cmidrule(lr){2-4} \cmidrule(lr){5-7} \cmidrule(lr){8-10}
         & {LSD $\downarrow$} & {SiSpec $\uparrow$} & {ViSQOL $\uparrow$} 
         & {LSD $\downarrow$} & {SiSpec $\uparrow$} & {ViSQOL $\uparrow$} 
         & {LSD $\downarrow$} & {SiSpec $\uparrow$} & {ViSQOL $\uparrow$} \\
        \midrule
        AudioSR & 
            \num{1.745} & \num{21.740} & \num{3.391} & 
            \num{1.809} & \num{27.259} & \num{3.226} &
            \num{1.854} & \num{28.974} & \num{3.150} \\
        CQTDiff & 
        \num{1.738} & \num{10.620} & \num{1.747} & 
        \num{1.631} & \num{17.417} & \num{1.777} & 
        \num{1.574} & \num{21.615} & \num{2.000} \\
        IBAR & 
        \bestnumber{\num{1.12}}  & \num{12.31} & \num{2.995} & 
        \bestnumber{\num{0.924}} & \num{12.94} & \num{3.525} & 
        \bestnumber{\num{0.849}} & \num{13.08} & \num{3.843} \\ \hline
        \rowcolor{pink!20} \modelname (no partitioning) & 
        \num{1.33}  & \num{25.51} & \num{2.557} & 
        \num{1.05}  & \num{33.10} & \num{3.201} & 
        \num{0.867} & \num{35.34} & \num{3.936} \\
        \rowcolor{pink!20} \modelname (2-partitioning) & 
        \num{1.29}  & \bestnumber{\num{28.15}} & \num{3.101} & 
        \num{1.07}  & \bestnumber{\num{34.36}} & \num{3.718} & 
        \num{0.878} & \num{35.97}              & \num{4.200} \\
        \rowcolor{pink!20} \modelname (4-partitioning) & 
        \num{1.77} & \num{27.56}              & \bestnumber{\num{3.446}}   & 
        \num{1.59} & \num{34.25}              & \bestnumber{\num{3.829}}   & 
        \num{1.51} & \bestnumber{\num{36.07}} & \bestnumber{\num{4.274} }  \\
        \bottomrule
    \end{tabular}
    \label{tab: MTD BWE main}
\end{table}

\begin{table}[!t]
    \centering
    \caption{Bandwidth extension results on Maestro (classical piano music with MIDI).}
    \tiny
    \begin{tabular}{l S S[round-precision=2] S S S[round-precision=2] S S S[round-precision=2] S}
        \toprule
        \multirow{2}{*}{\textbf{Method}} & \multicolumn{3}{c}{\textbf{Cutoff = 4kHz}} & \multicolumn{3}{c}{\textbf{Cutoff = 8kHz}} & \multicolumn{3}{c}{\textbf{Cutoff = 12kHz}} \\
        \cmidrule(lr){2-4} \cmidrule(lr){5-7} \cmidrule(lr){8-10}
         & {LSD $\downarrow$} &  {SiSpec $\uparrow$} & {$\text{F}_1$ $\uparrow$} & {LSD $\downarrow$} & {SiSpec $\uparrow$} & {$\text{F}_1$ $\uparrow$} & {LSD $\downarrow$} & {SiSpec $\uparrow$} & {$\text{F}_1$ $\uparrow$} \\
        \midrule
        CQTDiff & \num{1.154} & \num{31.486} & \num{0.761} & \num{1.137} & \num{32.993} & \num{0.772} & \num{1.129} & \num{33.334} & \num{0.774} \\
        IBAR & \bestnumber{\num{0.769}} & \num{12.687} & \num{0.769} & \num{0.688} & \num{12.220} & \num{0.757} & \num{0.616} & \num{13.475} & \num{0.770} \\ \hline
        \rowcolor{pink!20} \modelname (4-partitioning) & \num{0.773} & \bestnumber{\num{34.315}} & \bestnumber{\num{0.910}} & \bestnumber{\num{0.659}} & \bestnumber{\num{41.690}} & \bestnumber{\num{0.910}} & \bestnumber{\num{0.545}} & \bestnumber{\num{42.597}} & \bestnumber{\num{0.910}} \\
        \bottomrule
    \end{tabular}
    \label{tab: Maestro BWE main}
\end{table}

\newpage
\subsection{Inpainting Results}

The inpainting results are shown in Tables \ref{tab: AAM inpainting main} - \ref{tab: Maestro inpainting main}. \modelname achieves consistently better evaluation results. This is likely because inpainting has a smaller synthesized-content versus context ratio than bandwidth extension.
All three \modelname models achieve similar generation quality. Models with $t$-range partitioning have slightly better quality than the one without partitioning, and having 4 partitions is slightly better than 2. 

\begin{figure}[!t]
    \centering
    \includegraphics[width=0.8\linewidth]{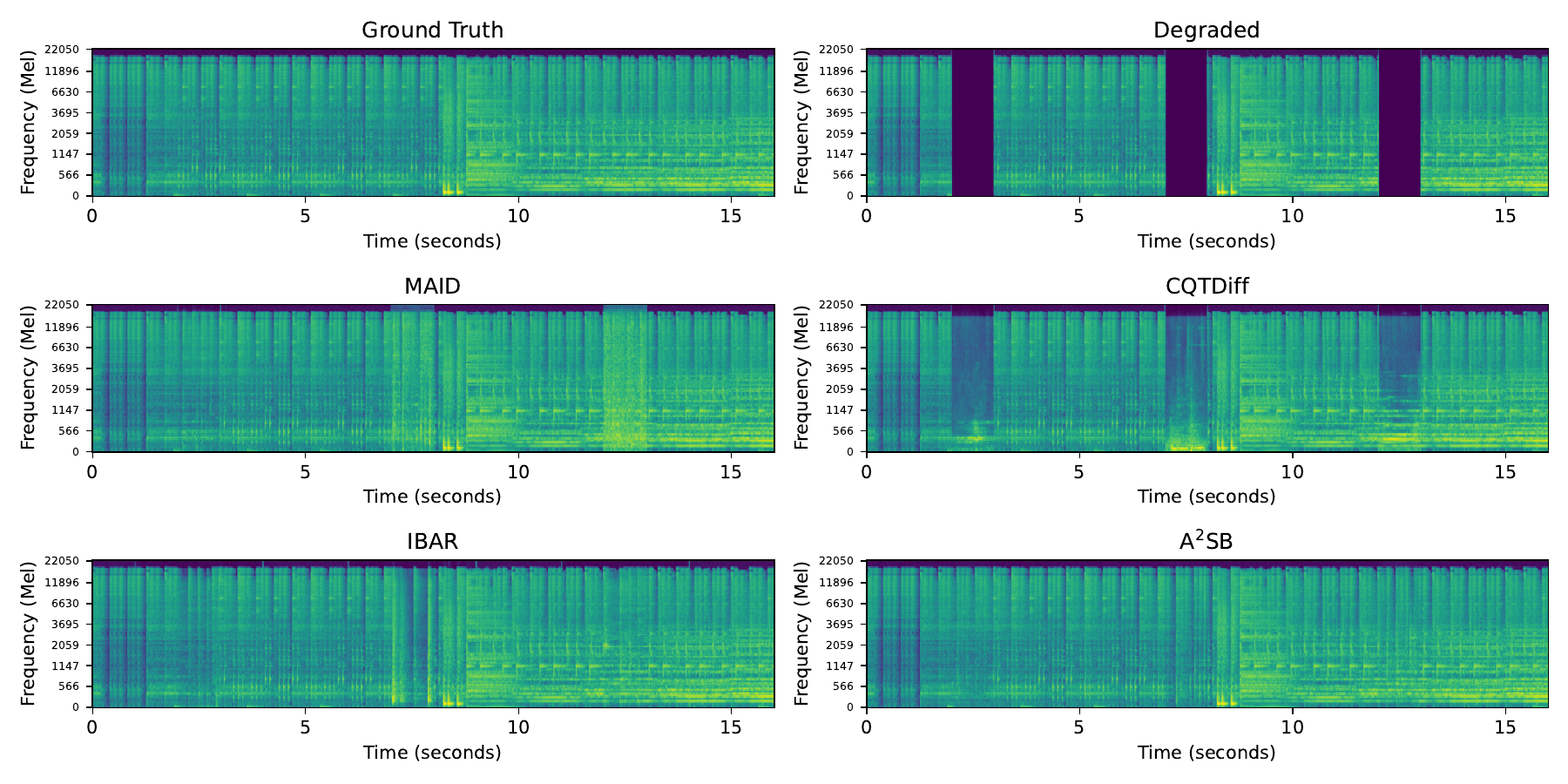}
    \caption{Qualitative comparison between different inpainting methods with inpainting gap = 1 sec.}
    \label{fig: inp example main}
\end{figure}

In Figure \ref{fig: inp example main}, we show a qualitative sample of different inpainting baselines. More samples can be found in Appendix \ref{app: INP samples}. MAID and CQTDiff have much worse quality, and \modelname has more consistent outputs compared to that of baselines. 

\begin{table}[!t]
    \centering
    \caption{Inpainting results on AAM (synthetic music).}
    \tiny
    \begin{tabular}{l|S S[round-precision=2] S S S[round-precision=2] S S S[round-precision=2] S}
        \toprule
        \multirow{2}{*}{\textbf{Method}} & 
        \multicolumn{3}{c}{\textbf{Gap = 300ms}} & 
        \multicolumn{3}{c}{\textbf{Gap = 500ms}} & 
        \multicolumn{3}{c}{\textbf{Gap = 1000ms}} \\
        \cmidrule(lr){2-4} \cmidrule(lr){5-7} \cmidrule(lr){8-10}
         & {LSD $\downarrow$} & {SiSpec $\uparrow$} & {ViSQOL $\uparrow$} 
         & {LSD $\downarrow$} & {SiSpec $\uparrow$} & {ViSQOL $\uparrow$} 
         & {LSD $\downarrow$} & {SiSpec $\uparrow$} & {ViSQOL $\uparrow$} \\
        \midrule
        MAID & 
        \num{0.139} & \num{14.370} & \num{4.570} & 
        \num{0.208} & \num{11.420} & \num{4.504} & 
        \num{0.378} & \num{7.740}  & \num{4.305} \\
        CQTDiff & 
        \num{1.516} & \num{14.365} & \num{4.502} & 
        \num{1.510} & \num{11.127} & \num{4.457} & 
        \num{1.494} & \num{7.168}  & \num{4.219} \\
        IBAR & 
        \num{0.512} & \num{8.670} & \num{4.231} & 
        \num{0.420} & \num{9.660} & \num{4.383} & 
        \num{0.525} & \num{6.880} & \num{4.204} \\ \hline
        \rowcolor{pink!20} \modelname (no partitioning) & 
        \num{0.081} & \num{17.100} & \num{4.660} & 
        \num{0.128} & \num{12.720} & \num{4.592} & 
        \num{0.257} & \num{7.900}  & \num{4.432} \\
        \rowcolor{pink!20} \modelname (2-partitioning) & 
        \num{0.077} & \num{17.890}                          & \num{4.666} & 
        \num{0.122} & \num{13.950}                          & \num{4.601} & 
        \bestnumber{\num{0.238}} & \bestnumber{\num{9.310}} & \num{4.442} \\
        \rowcolor{pink!20} \modelname (4-partitioning) & 
        \bestnumber{\num{0.076}} & \bestnumber{\num{18.360}} & \bestnumber{\num{4.673}} & 
        \bestnumber{\num{0.121}} & \bestnumber{\num{13.960}} & \bestnumber{\num{4.613}} & 
        \bestnumber{\num{0.238}} & \num{9.160}               & \bestnumber{\num{4.465}} \\
        \bottomrule
    \end{tabular}
    \label{tab: AAM inpainting main}
\end{table}

\begin{table}[!t]
    \centering
    \caption{Inpainting results on CCMixter (remixed music).}
    \tiny
    \begin{tabular}{l|S S[round-precision=2] S S S[round-precision=2] S S S[round-precision=2] S}
        \toprule
        \multirow{2}{*}{\textbf{Method}} & 
        \multicolumn{3}{c}{\textbf{Gap = 300ms}} & 
        \multicolumn{3}{c}{\textbf{Gap = 500ms}} & 
        \multicolumn{3}{c}{\textbf{Gap = 1000ms}} \\
        \cmidrule(lr){2-4} \cmidrule(lr){5-7} \cmidrule(lr){8-10}
         & {LSD $\downarrow$} & {SiSpec $\uparrow$} & {ViSQOL $\uparrow$} 
         & {LSD $\downarrow$} & {SiSpec $\uparrow$} & {ViSQOL $\uparrow$} 
         & {LSD $\downarrow$} & {SiSpec $\uparrow$} & {ViSQOL $\uparrow$} \\
        \midrule
        MAID & 
        \num{0.129} & \num{13.340} & \num{4.556} & 
        \num{0.205} & \num{10.670} & \num{4.462} & 
        \num{0.394} & \num{7.110}  & \num{4.235} \\
        CQTDiff & 
        \num{1.305} & \num{11.162} & \num{4.486} & 
        \num{1.293} & \num{9.011}  & \num{4.403} & 
        \num{1.266} & \num{5.954}  & \num{4.126} \\
        IBAR & 
        \num{0.384} & \num{10.890} & \num{4.466} & 
        \num{0.415} & \num{9.360}  & \num{4.378} & 
        \num{0.504} & \num{6.560}  & \num{4.186} \\ \hline
        \rowcolor{pink!20} \modelname (no partitioning) & 
        \num{0.088} & \num{13.830} & \num{4.625} & 
        \num{0.139} & \num{10.680} & \num{4.537} & 
        \num{0.274} & \num{6.610}  & \num{4.336} \\
        \rowcolor{pink!20} \modelname (2-partitioning) & 
        \bestnumber{\num{0.086}} & \bestnumber{\num{15.210}} & \num{4.630} & 
        \bestnumber{\num{0.134}} & \bestnumber{\num{12.310}} & \num{4.547} & 
        \bestnumber{\num{0.259}} & \bestnumber{\num{8.480}}  & \num{4.352} \\
        \rowcolor{pink!20} \modelname (4-partitioning) & 
        \bestnumber{\num{0.086}} & \num{14.890} & \bestnumber{\num{4.632}} & 
        \num{0.135} & \num{11.880}              & \bestnumber{\num{4.549}} & 
        \num{0.261} & \num{7.960}               & \bestnumber{\num{4.358}} \\
        \bottomrule
    \end{tabular}
    \label{tab: CCMixter inpainting main}
\end{table}

\begin{table}[!t]
    \centering
    \caption{Inpainting results on MTD (classical music).}
    \tiny
    
    \begin{tabular}{l|S S[round-precision=2] S S S[round-precision=2] S S S[round-precision=2] S}
        \toprule
        \multirow{2}{*}{\textbf{Method}} & 
        \multicolumn{3}{c}{\textbf{Gap = 300ms}} & 
        \multicolumn{3}{c}{\textbf{Gap = 500ms}} & 
        \multicolumn{3}{c}{\textbf{Gap = 1000ms}} \\
        \cmidrule(lr){2-4} \cmidrule(lr){5-7} \cmidrule(lr){8-10}
         & {LSD $\downarrow$} & {SiSpec $\uparrow$} & {ViSQOL $\uparrow$} 
         & {LSD $\downarrow$} & {SiSpec $\uparrow$} & {ViSQOL $\uparrow$} 
         & {LSD $\downarrow$} & {SiSpec $\uparrow$} & {ViSQOL $\uparrow$} \\
        \midrule
        MAID & 
        \num{0.139} & \num{9.860} & \num{4.406} &
        \num{0.223} & \num{7.290} & \num{4.285} &
        \num{0.430} & \num{3.790} & \num{4.044} \\
        CQTDiff & 
        \num{0.846} & \num{8.844} & \num{4.411}& 
        \num{0.855} & \num{5.817} & \num{4.252}& 
        \num{0.877} & \num{1.263} & \num{3.963}\\
        IBAR & 
        \num{0.293} & \num{13.620} & \num{4.136} & 
        \num{0.306} & \num{11.710} & \num{4.109} & 
        \num{0.346} & \num{7.930}  & \num{4.030} \\ \hline
        \rowcolor{pink!20} \modelname (no partitioning) & 
        \num{0.073} & \num{17.830} & \num{4.641} & 
        \num{0.106} & \num{13.870} & \num{4.562} & 
        \num{0.201} & \num{7.800}  & \num{4.347} \\
        \rowcolor{pink!20} \modelname (2-partitioning) & 
        \bestnumber{\num{0.071}} & \num{18.280}              & \num{4.650} & 
        \bestnumber{\num{0.103}} & \bestnumber{\num{14.740}} & \num{4.572} & 
        \bestnumber{\num{0.187}} & \bestnumber{\num{9.940}}  & \num{4.376} \\
        \rowcolor{pink!20} \modelname (4-partitioning) & 
        \bestnumber{\num{0.071}} & \bestnumber{\num{18.430}} & \bestnumber{\num{4.655}} & 
        \bestnumber{\num{0.103}} & \num{14.730}              & \bestnumber{\num{4.584}} & 
        \bestnumber{\num{0.187}} & \num{9.400}               & \bestnumber{\num{4.402}} \\
        \bottomrule
    \end{tabular}
    \label{tab: MTD inpainting main}
\end{table}

\begin{table}[!t]
    \centering
    \caption{Inpainting results on Maestro (classical piano music with MIDI).}
    \tiny
    \begin{tabular}{l S S[round-precision=2] S S S[round-precision=2] S S S[round-precision=2] S}
        \toprule
        \multirow{2}{*}{\textbf{Method}} & \multicolumn{3}{c}{\textbf{Gap = 300ms}} & \multicolumn{3}{c}{\textbf{Gap = 500ms}} & \multicolumn{3}{c}{\textbf{Gap = 1000ms}} \\
        \cmidrule(lr){2-4} \cmidrule(lr){5-7} \cmidrule(lr){8-10}
         & {LSD $\downarrow$} &  {SiSpec $\uparrow$} & {$\text{F}_1$ $\uparrow$} & {LSD $\downarrow$} & {SiSpec $\uparrow$} & {$\text{F}_1$ $\uparrow$} & {LSD $\downarrow$} & {SiSpec $\uparrow$} & {$\text{F}_1$ $\uparrow$} \\
        \midrule
        MAID & \num{0.700} & \num{8.404} & \num{0.673} & \num{0.831} & \num{6.163} & \num{0.666} & \num{1.156} & \num{2.897} & \num{0.655} \\
        CQTDiff & \num{0.691} & \num{12.243} & \num{0.818} & \num{0.703} & \num{8.525} & \num{0.814} & \num{0.741} & \num{4.382} & \num{0.798} \\
        IBAR & \num{0.344} & \num{12.728} & \num{0.803} & \num{0.381} & \num{9.503} & \num{0.795} & \num{0.413} & \num{6.284} & \num{0.786} \\ \hline
        \rowcolor{pink!20} \modelname (4-partitioning) & \bestnumber{\num{0.134}} & \bestnumber{\num{17.033}} & \bestnumber{\num{0.870}} & \bestnumber{\num{0.167}} & \bestnumber{\num{13.326}} & \bestnumber{\num{0.854}} & \bestnumber{\num{0.254}} & \bestnumber{\num{8.446}} & \bestnumber{\num{0.820}} \\
        \bottomrule
    \end{tabular}
    \label{tab: Maestro inpainting main}
\end{table}

~~\newpage
\subsection{Human Evaluation}

We additionally conduct human evaluation on the bandwidth extension (cutoff = 4kHz) and inpainting (gap = 1000ms) experiments. For each OOD test dataset, we randomly select fifty segments and ask human listeners to rate the output quality based on how close they sound compared to the ground truth sample. The results are shown in Tables \ref{tab: MOS bwe main} and \ref{tab: MOS inpainting main}. We find {\modelname consistently and significantly outperforms baselines on both tasks and all three test sets}. We also find having more partitions consistently improves subjective scores. Especially, the improvements from no partitioning to 2-partitioning are solid, and further increasing the number of partitions to 4 yields diminishing returns. Combining all objective and subjective results, we conclude that \textit{\textcolor{forestgreen}{the 4-partitioned \modelname has the best overall quality}}, but \textit{\textcolor{forestgreen}{the 2-partitioned \modelname constitutes a better cost-performance ratio}}.

\begin{table}[!h]
    \centering
    \caption{Human evaluation results on bandwidth extension.}
    \scriptsize
    \begin{tabular}{l|ccc}
        \toprule
        \multirow{2}{*}{\textbf{Method}} & 
        \multicolumn{3}{c}{\textbf{MOS (bandwidth extension)}} \\
        \cmidrule(lr){2-4} & {AAM} & {CCMixter} & {MTD} \\ \midrule
        Ground Truth & 
        $4.36 \pm 0.04$ & $4.39 \pm 0.05$ & $4.26 \pm 0.04$ \\ \midrule
        AudioSR & 
        $3.65 \pm 0.08$ & $3.67 \pm 0.08$ & $3.72 \pm 0.10$ \\
        CQTDiff & 
        $3.85 \pm 0.08$ & $3.10 \pm 0.12$ & $2.99 \pm 0.10$ \\
        IBAR & 
        $3.89 \pm 0.07$ & $2.96 \pm 0.13$ & $3.75 \pm 0.07$ \\ \hline
        \rowcolor{pink!20} \modelname (no partitioning) & 
        $4.13 \pm 0.06$ & $4.10 \pm 0.07$ & $3.79 \pm 0.07$ \\
        \rowcolor{pink!20} \modelname (2-partitioning) & 
        \bestnumber{$4.17 \pm 0.06$} & \bestnumber{$4.17 \pm 0.06$} & \bestnumber{$3.96 \pm 0.06$} \\
        \rowcolor{pink!20} \modelname (4-partitioning) & 
        \bestnumber{$4.18 \pm 0.06$} & $4.08 \pm 0.06$ & \bestnumber{$3.97 \pm 0.06$} \\
        \bottomrule
    \end{tabular}
    \label{tab: MOS bwe main}
\end{table}

\begin{table}[!h]
    \centering
    \caption{Human evaluation results on inpainting.}
    \scriptsize
    \begin{tabular}{l|ccc}
        \toprule
        \multirow{2}{*}{\textbf{Method}} & 
        \multicolumn{3}{c}{\textbf{MOS (inpainting)}} \\
        \cmidrule(lr){2-4} & {AAM} & {CCMixter} & {MTD} \\ \midrule
        Ground Truth & 
        $4.41 \pm 0.05$ & $4.36 \pm 0.04$ & $4.38 \pm 0.05$ \\ \midrule
        MAID & 
        $3.27 \pm 0.10$ & $3.28 \pm 0.10$ & $3.33 \pm 0.10$ \\
        CQTDiff & 
        $3.59 \pm 0.08$ & $3.64 \pm 0.09$ & $3.63 \pm 0.09$ \\
        IBAR & 
        $3.70 \pm 0.08$ & $3.69 \pm 0.08$ & $3.96 \pm 0.07$ \\ \hline
        \rowcolor{pink!20} \modelname (no partitioning) & 
        $3.96 \pm 0.07$ & $3.82 \pm 0.08$ & $4.02 \pm 0.07$ \\
        \rowcolor{pink!20} \modelname (2-partitioning) & 
        $4.00 \pm 0.07$ & $3.85 \pm 0.08$ & \bestnumber{$4.09 \pm 0.06$} \\
        \rowcolor{pink!20} \modelname (4-partitioning) & 
        \bestnumber{$4.06 \pm 0.06$} & \bestnumber{$3.92 \pm 0.07$} & \bestnumber{$4.10 \pm 0.06$} \\
        \bottomrule
    \end{tabular}
    \label{tab: MOS inpainting main}
\end{table}

We then investigate how accurately the objective metrics could predict perceptual quality (MOS). We compute the Spearman Correlation between MOS and each objective metric ($-$LSD, SiSpec, and ViSQOL) in Table \ref{tab: MOS Spearman Correlation}. Results indicate all the three objective metrics are moderately correlated with the MOS metric, but far from perfect. \footnote{
We additionally fit linear regression between MOS and objective metrics, and obtain the following results. For bandwidth extension, 
$
    \mathrm{MOS}=3.2158-0.0411\times\mathrm{LSD}+0.1567\times\mathrm{sign}(\mathrm{SiSpec})\times\log|\mathrm{SiSpec}|+0.1015\times\mathrm{ViSQOL} ~~ (R^2=0.311).
$
For inpainting, 
$
    \mathrm{MOS}=4.1775-1.4022\times\mathrm{LSD}+0.0681\times\mathrm{sign}(\mathrm{SiSpec})\times\log|\mathrm{SiSpec}|+0.0649\times\mathrm{ViSQOL} ~~ (R^2=0.252).
$
}

\begin{table}[!h]
    \centering
    \small
    \caption{Spearman Correlation between MOS and objective metrics. All p-values are less than $0.001$.}
    \begin{tabular}{l|S S S}
        \toprule
        \textbf{Task} & \textbf{$-$LSD} & \textbf{SiSpec} & \textbf{ViSQOL} \\ \midrule
        {Bandwidth extension (cutoff = 4kHz)} & \num{0.443} & \num{0.491} & \num{0.450} \\
        {Inpainting (gap = 1000ms)} & \num{0.549} & \num{0.461} & \num{0.480} \\
        \bottomrule
    \end{tabular}
    \label{tab: MOS Spearman Correlation}
\end{table}

\subsection{Necessity of Factorized Audio Representation}
\label{sec: experiments: factorization}

In this section, we demonstrate that compared to the simple complex representation ($S$ in Section \ref{sec: method: magphase}), our three-channel factorized audio representation \eqref{eq: audio representation} leads to better fit of the magnitude spectrogram. This thus confirms that factorizing magnitude out from phase and treating them as separate channels prevents the difficulties of modeling phase values from affecting the magnitude modeling task.

\begin{figure}[!t]
    \centering
    \begin{subfigure}{0.8\linewidth}
        \centering
        \includegraphics[width=\linewidth]{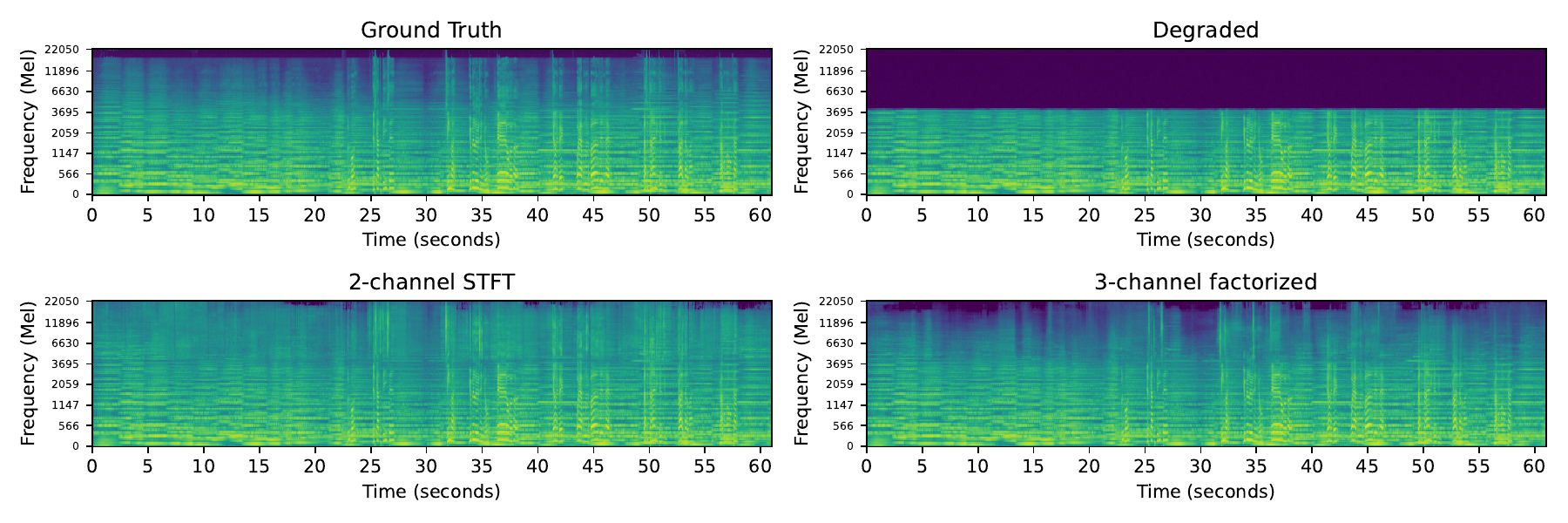}
        \caption{Comparing audio representations: sample 1.}
    \end{subfigure}
    \hfill
    \begin{subfigure}{0.8\linewidth}
        \centering
        \includegraphics[width=\linewidth]{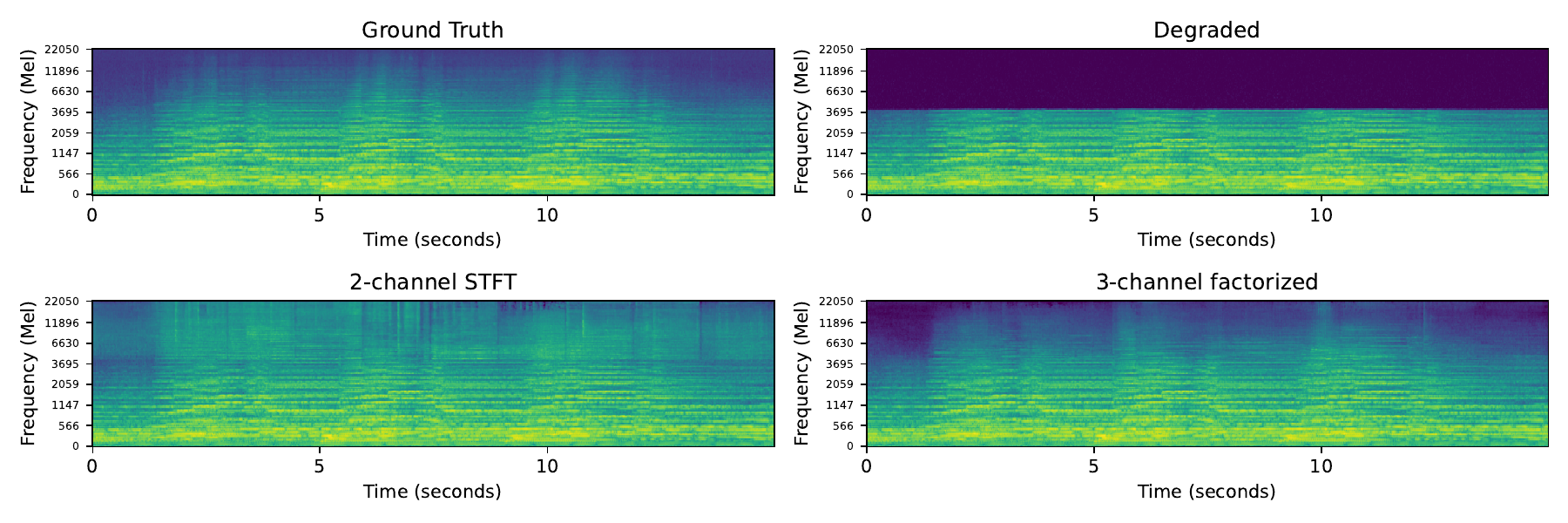}
        \caption{Comparing audio representations: sample 2.}
    \end{subfigure}
    \caption{Qualitative comparison between \modelname trained with two-channel STFT representation ($S$) and our proposed three-channel factorized representation \eqref{eq: audio representation}. The model trained with the two-channel STFT representation has artifacts around the cutoff frequency and predicts too much content for higher frequencies, validating the effectiveness of our three-channel factorized representation.}
    \label{fig: mag_2D_vs_3D_comparison}
\end{figure}

In Figure \ref{fig: mag_2D_vs_3D_comparison}, we visualize generated samples of our model trained on STFT and factorized representations, respectively. It is clearly seen that the STFT representation leads to artifacts around the cutoff frequency, while this is alleviated in our factorized audio representation. The overestimation of higher frequency magnitudes is also visible in the 2-channel spectrogram. These results suggest that using the factorized representation is often a better choice and should have more stable learning dynamics.

In Figure \ref{fig: mag_decay_comparison}, we report the average magnitude at different frequency bands. In detail, we randomly select 10 music samples from each test set for this experiment, and report the averaged magnitude over these samples. Results indicate that the complex representation poorly estimates magnitude in all frequency bands. In contrast, our three-channel factorized representation leads to similar magnitude mass compared to ground truth.

\begin{figure}[!h]
    \centering
    \begin{tikzpicture}

\begin{groupplot}[
    group style={
        group size=1 by 3, 
        vertical sep=1.5cm   
    },
    width=10cm,
    height=4cm,
    title={AAM},
    xlabel={Frequency Range (kHz)},
    ylabel={Average Magnitude},
    legend style={
        at={(0.99,0.97)},
        anchor=north east,
        legend columns=1,
        font=\scriptsize
    },
    grid=both,
    grid style={dashed, gray!30},
    xticklabels={dc, dc,$4\text{-}6$, $6\text{-}8$, $8\text{-}10$, $10\text{-}12$, $12\text{-}14$, $14\text{-}16$, $16\text{-}18$, $18\text{-}20$, $20\text{-}22$}, 
    ymin=0, ymax=.8,
]

\pgfplotstableread[col sep=comma]{magnitude_analysis.csv}{\datatable}
\nextgroupplot[
    title={Average magnitude of output spectrogram on AAM},
    title style={at={(0.5,1.1)}, anchor=north},
]
\addplot[color=CustomBlue,mark=triangle*,line width=0.8pt] table[x=freq_start, y=aam_gt] {\datatable};
\addlegendentry{Ground truth}
\addplot[color=CustomRed,mark=square*,line width=0.8pt] table[x=freq_start, y=aam_mip] {\datatable};
\addlegendentry{\modelname (3-channel factorized)}
\addplot[color=forestgreen,mark=pentagon*,line width=0.8pt] table[x=freq_start, y=aam_complex] {\datatable};
\addlegendentry{\modelname (2-channel STFT)}

\nextgroupplot[
    title={Average magnitude of output spectrogram on CCMixter},
    title style={at={(0.5,1.1)}, anchor=north},
    ymax=1.5
]
\addplot[color=CustomBlue,mark=triangle*,line width=0.8pt] table[x=freq_start, y=ccmixter_gt] {\datatable};
\addlegendentry{Ground truth}
\addplot[color=CustomRed,mark=square*,line width=0.8pt] table[x=freq_start, y=ccmixter_mip] {\datatable};
\addlegendentry{\modelname (3-channel factorized)}
\addplot[color=forestgreen,mark=pentagon*,line width=0.8pt] table[x=freq_start, y=ccmixter_complex] {\datatable};
\addlegendentry{\modelname (2-channel STFT)}

\nextgroupplot[
    title={Average magnitude of output spectrogram on MTD},
    title style={at={(0.5,1.1)}, anchor=north},
    ymax=.1
]
\addplot[color=CustomBlue,mark=triangle*,line width=0.8pt] table[x=freq_start, y=mtd_gt] {\datatable};
\addlegendentry{Ground truth}
\addplot[color=CustomRed,mark=square*,line width=0.8pt] table[x=freq_start, y=mtd_mip] {\datatable};
\addlegendentry{\modelname (3-channel factorized)}
\addplot[color=forestgreen,mark=pentagon*,line width=0.8pt] table[x=freq_start, y=mtd_complex] {\datatable};
\addlegendentry{\modelname (2-channel STFT)}

\end{groupplot}
\end{tikzpicture}
    \vspace{-2em}
    \caption{This plot compares the average spectrogram magnitude of outputs from models trained with different audio representations: 2-channel STFT and our proposed 3-channel factorized representation \eqref{eq: audio representation}. The results in this plot demonstrate that jointly modeling phase and magnitude without uncoupling may result in overall inaccurate magnitude generations compared to that of the target distribution.}
    \label{fig: mag_decay_comparison}
\end{figure}

\subsection{Necessity of Phase Orthogonalization}

\begin{figure}[!h]
    \centering
    \begin{subfigure}[t]{0.34\textwidth}      \begin{tikzpicture}
\begin{axis}[
    ytick={1, 2, 3, 4, 5, 6, 7, 8, 9},
    yticklabels={4-6 kHz, 6-8 kHz, 8-10 kHz, 10-12 kHz, 12-14 kHz, 14-16 kHz, 16-18 kHz, 18-20 kHz, 20-22 kHz},
    enlargelimits=.15,
    axis lines=left,
    title={AAM},
    width=4.5cm,
    height=4.5cm,
    xmode=log,
    ymin=0,
    xlabel={$\mathbf{Err}_{\text{phase-ortho}}$},
    yticklabel style={font=\tiny}
]

\addplot+[
    draw=CustomBlue,
    solid,
    boxplot prepared={
        lower quartile=0.000002, upper quartile=0.000031,
        median=0.000012, 
         upper whisker=0.001953,
        box extend=0.25
    }
] coordinates {};

\addplot+[
    draw=CustomBlue,
    solid,
    boxplot prepared={
        lower quartile=0.000002, upper quartile=0.000031,
        median=0.000012, 
         upper whisker=0.001833,
        box extend=0.25
    }
] coordinates {};

\addplot+[
    draw=CustomBlue,
    solid,
    boxplot prepared={
        lower quartile=0.000004, upper quartile=0.000039,
        median=0.000017, 
         upper whisker=0.002303,
        box extend=0.25
    }
] coordinates {};

\addplot+[
    draw=CustomBlue,
    solid,
    boxplot prepared={
        lower quartile=0.000002, upper quartile=0.000031,
        median=0.000012, 
         upper whisker=0.001442,
        box extend=0.25
    }
] coordinates {};

\addplot+[
    draw=CustomBlue,
    solid,
    boxplot prepared={
        lower quartile=0.000002, upper quartile=0.000023,
        median=0.000008, 
         upper whisker=0.001468,
        box extend=0.25
    }
] coordinates {};

\addplot+[
    draw=CustomBlue,
    solid,
    boxplot prepared={
        lower quartile=0.000002, upper quartile=0.000031,
        median=0.000008, 
         upper whisker=0.001549,
        box extend=0.25
    }
] coordinates {};

\addplot+[
    draw=CustomBlue,
    solid,
    boxplot prepared={
        lower quartile=0.000002, upper quartile=0.000031,
        median=0.000012, 
         upper whisker=0.001495,
        box extend=0.25
    }
] coordinates {};

\addplot+[
    draw=CustomBlue,
    solid,
    boxplot prepared={
        lower quartile=0.000002, upper quartile=0.000031,
        median=0.000012, 
         upper whisker=0.001289,
        box extend=0.25
    }
] coordinates {};

\addplot+[
    draw=CustomBlue,
    solid,
    boxplot prepared={
        lower quartile=0.000002, upper quartile=0.000017,
        median=0.000008, 
         upper whisker=0.000322,
        box extend=0.25
    }
] coordinates {};

\end{axis}
\end{tikzpicture}
    \end{subfigure}
    \begin{subfigure}[t]{0.3\textwidth}
    \begin{tikzpicture}
\begin{axis}[
    ytick={1, 2, 3, 4, 5, 6, 7, 8, 9},
    yticklabels={},
    enlargelimits=.15,
    axis lines=left,
    title={MTD},
    width=5cm,
    height=4.5cm,
    xmode=log,
    ymin=0,
    xlabel={$\mathbf{Err}_{\text{phase-ortho}}$},
    yticklabel style={font=\tiny}
]

\addplot+[
    draw=CustomBlue,
    solid,
    boxplot prepared={
        lower quartile=0.000002, upper quartile=0.000023,
        median=0.000008, 
         upper whisker=0.00279,
        box extend=0.25
    }
] coordinates {};

\addplot+[
    draw=CustomBlue,
    solid,
    boxplot prepared={
        lower quartile=0.000002, upper quartile=0.000031,
        median=0.000012, 
         upper whisker=0.001633,
        box extend=0.25
    }
] coordinates {};

\addplot+[
    draw=CustomBlue,
    solid,
    boxplot prepared={
        lower quartile=0.000002, upper quartile=0.000031,
        median=0.000012, 
         upper whisker=0.001774,
        box extend=0.25
    }
] coordinates {};

\addplot+[
    draw=CustomBlue,
    solid,
    boxplot prepared={
        lower quartile=0.000002, upper quartile=0.000031,
        median=0.000012, 
         upper whisker=0.000881,
        box extend=0.25
    }
] coordinates {};

\addplot+[
    draw=CustomBlue,
    solid,
    boxplot prepared={
        lower quartile=0.000002, upper quartile=0.000023,
        median=0.000008, 
         upper whisker=0.000567,
        box extend=0.25
    }
] coordinates {};

\addplot+[
    draw=CustomBlue,
    solid,
    boxplot prepared={
        lower quartile=0.000002, upper quartile=0.000031,
        median=0.000012, 
         upper whisker=0.000488,
        box extend=0.25
    }
] coordinates {};

\addplot+[
    draw=CustomBlue,
    solid,
    boxplot prepared={
        lower quartile=0.000002, upper quartile=0.000031,
        median=0.000012, 
         upper whisker=0.000763,
        box extend=0.25
    }
] coordinates {};

\addplot+[
    draw=CustomBlue,
    solid,
    boxplot prepared={
        lower quartile=0.000002, upper quartile=0.000023,
        median=0.000008, 
         upper whisker=0.000601,
        box extend=0.25
    }
] coordinates {};

\addplot+[
    draw=CustomBlue,
    solid,
    boxplot prepared={
        lower quartile=0.000002, upper quartile=0.000017,
        median=0.000004, 
         upper whisker=0.0002,
        box extend=0.25
    }
] coordinates {};

\end{axis}
\end{tikzpicture}
    \end{subfigure}
    \begin{subfigure}[t]{0.3\textwidth}
    \begin{tikzpicture}
\begin{axis}[
    ytick={1, 2, 3, 4, 5, 6, 7, 8, 9},
    yticklabels={},
    enlargelimits=.15,
    axis lines=left,
    title={CCMixter},
    width=5cm,
    height=4.5cm,
    xmode=log,
    ymin=0,
    xlabel={$\mathbf{Err}_{\text{phase-ortho}}$},
    yticklabel style={font=\tiny}
]

\addplot+[
    draw=CustomBlue,
    solid,
    boxplot prepared={
        lower quartile=0.000002, upper quartile=0.000031,
        median=0.000012, 
         upper whisker=0.002682,
        box extend=0.25
    }
] coordinates {};

\addplot+[
    draw=CustomBlue,
    solid,
    boxplot prepared={
        lower quartile=0.000002, upper quartile=0.000031,
        median=0.000012, 
         upper whisker=0.002357,
        box extend=0.25
    }
] coordinates {};

\addplot+[
    draw=CustomBlue,
    solid,
    boxplot prepared={
        lower quartile=0.000004, upper quartile=0.000039,
        median=0.000017, 
         upper whisker=0.002827,
        box extend=0.25
    }
] coordinates {};

\addplot+[
    draw=CustomBlue,
    solid,
    boxplot prepared={
        lower quartile=0.000004, upper quartile=0.000039,
        median=0.000017, 
         upper whisker=0.002140,
        box extend=0.25
    }
] coordinates {};

\addplot+[
    draw=CustomBlue,
    solid,
    boxplot prepared={
        lower quartile=0.000002, upper quartile=0.000031,
        median=0.000012, 
         upper whisker=0.001833,
        box extend=0.25
    }
] coordinates {};

\addplot+[
    draw=CustomBlue,
    solid,
    boxplot prepared={
        lower quartile=0.000002, upper quartile=0.000031,
        median=0.000008, 
         upper whisker=0.001604,
        box extend=0.25
    }
] coordinates {};

\addplot+[
    draw=CustomBlue,
    solid,
    boxplot prepared={
        lower quartile=0.000002, upper quartile=0.000031,
        median=0.000012, 
         upper whisker=0.001458,
        box extend=0.25
    }
] coordinates {};

\addplot+[
    draw=CustomBlue,
    solid,
    boxplot prepared={
        lower quartile=0.000002, upper quartile=0.000023,
        median=0.000008, 
         upper whisker=0.001406,
        box extend=0.25
    }
] coordinates {};

\addplot+[
    draw=CustomBlue,
    solid,
    boxplot prepared={
        lower quartile=0.000002, upper quartile=0.000008,
        median=0.000004, 
         upper whisker=0.000231,
        box extend=0.25
    }
] coordinates {};

\end{axis}
\end{tikzpicture}
    \end{subfigure}
    \caption{These box-plots visualize the distribution of the ($\log$-scale) phase orthogonalization error $\mathbf{Err}_{\text{phase-ortho}}$ in \eqref{eq: orthogonalization error} without any orthogonalization correction. The left-most whisker is omitted and is effectively zero. The right most whisker represents the $99.9$-th percentile, where the outlying $.01\%$ is omitted from the graph and may have an error of as high as $1.5$. The results indicate that our model can predict very accurate trigonometric values of phase for most of the time, and the phase orthogonalization acts primarily as an occasionally necessary safeguard.}
    \label{fig:svd-error}
\end{figure}

We study the impact of applying phase orthogonalization in \eqref{eq: orthogonalization}, where we find that the model's output are sufficiently close to being proper rotations and require only small adjustments. In Figure~\ref{fig:svd-error}, we visualize the distribution of the phase orthogonalization error $\mathbf{Err}_{\text{phase-ortho}}(X_{i,j})$ in \eqref{eq: orthogonalization error} at different frequency bands. In detail, we consider the bandwidth extension task with cutoff = 4kHz. We take the generated part (above 4kHz) of the output spectrogram and uniformly split it into 9 bins along the frequency axis. We then plot the distribution of $\mathbf{Err}_{\text{phase-ortho}}$ values within each bin.

We note that orthogonalization error is very small (the average error is around the order of $10^{-5}$), indicating that our model is able to learn the proposed audio representation very well. Only a small fraction ($<.1\%$) of the spectrogram may have larger phase orthogonalization error (up to 1.5), which will be corrected by phase orthogonalization \eqref{eq: orthogonalization}. Overall, the phase orthogonalization provides the necessary guarantee to ensure proper STFT inversion, while likely having nominal impact on perceptual quality given the scale of its adjustments.

\newpage
\subsection{Memory Usage for Long Audio Restoration}

We study the GPU memory usage with MultiDiffusion enabled in our model. We consider the bandwidth extension experiment with a cutoff frequency of 4kHz, and use the no-partitioning model to record GPU memory usage. \footnote{Note that for partitioned models, we could move unused checkpoints to CPU for each $t$-range.} We demonstrate the results versus input audio length in Figure \ref{fig: multidiffusion memory plot}. The slope shows the memory usage from the cached vector fields in MultiDiffusion, which could be further optimized by moving them to the CPU after computing the vector field for each patch $\mathbf{int}_k$. The results indicate that our model can up-sample several minutes of audio on a common gaming GPU with $\sim10$G memory and over an hour on a professional GPU with $>50$G memory. We may obtain more memory reduction as well as acceleration by using TensorRT \footnote{\url{https://github.com/NVIDIA/TensorRT}} and custom CUDA kernels \footnote{\url{https://pytorch.org/tutorials/advanced/custom_ops_landing_page.html\#custom-ops-landing-page}}. 

\begin{figure}[!h]
    \centering
    \begin{tikzpicture}
    \begin{axis}[
        xlabel={Input Audio Length (Minutes)},
        ylabel={GPU Memory (GB)},
        xmin=0, xmax=60,
        ymin=0, ymax=60,
        xtick={0, 3, 6, 9, 15, 21, 30, 45, 60},
        ytick={0, 10, 20, 30, 40, 50, 60},
        width=12cm, 
        height=5cm,
        legend style={
            at={(0.98,0.05)},
            anchor=south east,
        },
    ]

    \addplot[color=CustomBlue,mark=triangle*,line width=0.8pt] coordinates {(0.25,6.86)(0.5,7.02)(0.75,7.47)(1,7.63)(1.5,8.21)(2,8.81)(3,9.86)(6,11.01)(9,12.75)(15,16.67)(21,20.33)(30,25.66)(45,35.67)(60,45.82)};
    
    \end{axis}

    \begin{axis}[
    at={(0.8cm,1.5cm)}, 
    anchor=south west,
    width=4cm, height=3cm,
    grid=both,
    xmin=0, xmax=3.25, 
    ymin=4, ymax=12, 
    ytick={5, 10},
    title={Zoomed-in plot},
    title style={font=\scriptsize, yshift=-2ex}
    ]
    \addplot[color=CustomBlue,mark=triangle*,line width=0.8pt] coordinates {(0.25,6.86)(0.5,7.02)(0.75,7.47)(1,7.63)(1.5,8.21)(2,8.81)(3,9.86)(6,11.01)(9,12.75)(15,16.67)(21,20.33)(30,25.66)(45,35.67)(60,45.82)};
    
    \end{axis}

    \end{tikzpicture}

    \caption{GPU memory usage verses input audio length (in minutes) at inference time with MultiDiffusion enabled. The GPU memory is recorded for the bandwidth extension experiment with a cutoff frequency at 4kHz. The results show that our model can up-sample several minutes of audio on a common gaming GPU and over an hour on a professional GPU.}
    \label{fig: multidiffusion memory plot}
\end{figure}

\subsection{The Effect of Sampling Steps}

We study the effect of sampling steps in \modelname. We conduct this experiment on a subset of the MTD dataset, and consider bandwidth extension with cutoff = 4kHz and inpainting with gap = 1000ms. The model is the 2-partitioning \modelname. Results are shown in Figures \ref{fig: BWE sampling steps} and \ref{fig: Inpainting sampling steps}. The results indicate that \modelname yields almost identical generation quality with different number of sampling steps as low as 25. 

\begin{figure}[!h]
    \centering
    
    \begin{minipage}[t]{0.32\textwidth}
    \begin{tikzpicture}
    \begin{axis}[
        xlabel={Number of sampling steps},
        xlabel style={font=\scriptsize,yshift=4pt},
        ylabel={LSD $\downarrow$},
        ylabel style={font=\scriptsize,yshift=-4pt},
        xmin=0, xmax=225,
        ymin=0.0, ymax=2.0,
        xtick={0,25,50,100,200},
        xticklabel style={font=\scriptsize},
        ytick={0.0,1.0,2.0},
        yticklabel style={font=\scriptsize},
        width=6cm,
    ]

    \addplot[color=CustomRed, mark=triangle*, line width=0.8pt] coordinates {(25,1.242)(50,1.259)(100,1.266)(200,1.270)};

    \end{axis}
    \end{tikzpicture}
    \end{minipage}
    \begin{minipage}[t]{0.32\textwidth}
    \begin{tikzpicture}
    \begin{axis}[
        xlabel={Number of sampling steps},
        xlabel style={font=\scriptsize,yshift=4pt},
        ylabel={SiSpec $\uparrow$},
        ylabel style={font=\scriptsize,yshift=-4pt},
        xmin=0, xmax=225,
        ymin=20.0, ymax=30.0,
        xtick={0,25,50,100,200},
        xticklabel style={font=\scriptsize},
        ytick={20,25,30},
        yticklabel style={font=\scriptsize},
        width=6cm,
    ]

    \addplot[color=CustomRed, mark=triangle*, line width=0.8pt] coordinates {(25,27.471)(50,27.423)(100,27.425)(200,27.427)};
    
    \end{axis}
    \end{tikzpicture}
    \end{minipage}
    \begin{minipage}[t]{0.32\textwidth}
    \begin{tikzpicture}
    \begin{axis}[
        xlabel={Number of sampling steps},
        xlabel style={font=\scriptsize,yshift=4pt},
        ylabel={ViSQOL $\uparrow$},
        ylabel style={font=\scriptsize,yshift=-4pt},
        xmin=0, xmax=225,
        ymin=0, ymax=5,
        xtick={0,25,50,100,200},
        xticklabel style={font=\scriptsize},
        ytick={0,1,2,3,4,5},
        yticklabel style={font=\scriptsize},
        width=6cm,
    ]

    \addplot[color=CustomRed, mark=triangle*, line width=0.8pt] coordinates {(25,3.058)(50,3.039)(100,3.001)(200,2.988)};
    
    \end{axis}
    \end{tikzpicture}
    \end{minipage}

    \caption{Objective evaluation results with different sampling steps in \modelname. We evaluate on a subset of MTD with cutoff = 4kHz. Results indicate \modelname has almost identical quality when we use less sampling steps.}
    
    \label{fig: BWE sampling steps}
\end{figure}

\begin{figure}[!h]
    \centering
    
    \begin{minipage}[t]{0.32\textwidth}
    \begin{tikzpicture}
    \begin{axis}[
        xlabel={Number of sampling steps},
        xlabel style={font=\scriptsize,yshift=4pt},
        ylabel={LSD $\downarrow$},
        ylabel style={font=\scriptsize,yshift=-4pt},
        xmin=0, xmax=225,
        ymin=0.0, ymax=1.0,
        xtick={0,25,50,100,200},
        xticklabel style={font=\scriptsize},
        ytick={0.0,0.5,1.0},
        yticklabel style={font=\scriptsize},
        width=6cm,
    ]

    \addplot[color=CustomRed, mark=triangle*, line width=0.8pt] coordinates {(25,0.188)(50,0.188)(100,0.189)(200,0.190)};

    \end{axis}
    \end{tikzpicture}
    \end{minipage}
    \begin{minipage}[t]{0.32\textwidth}
    \begin{tikzpicture}
    \begin{axis}[
        xlabel={Number of sampling steps},
        xlabel style={font=\scriptsize,yshift=4pt},
        ylabel={SiSpec $\uparrow$},
        ylabel style={font=\scriptsize,yshift=-4pt},
        xmin=0, xmax=225,
        ymin=0.0, ymax=20.0,
        xtick={0,25,50,100,200},
        xticklabel style={font=\scriptsize},
        ytick={0,10,20},
        yticklabel style={font=\scriptsize},
        width=6cm,
    ]

    \addplot[color=CustomRed, mark=triangle*, line width=0.8pt] coordinates {(25,9.917)(50,9.927)(100,9.852)(200,9.789)};
    
    \end{axis}
    \end{tikzpicture}
    \end{minipage}
    \begin{minipage}[t]{0.32\textwidth}
    \begin{tikzpicture}
    \begin{axis}[
        xlabel={Number of sampling steps},
        xlabel style={font=\scriptsize,yshift=4pt},
        ylabel={ViSQOL $\uparrow$},
        ylabel style={font=\scriptsize,yshift=-4pt},
        xmin=0, xmax=225,
        ymin=0, ymax=5,
        xtick={0,25,50,100,200},
        xticklabel style={font=\scriptsize},
        ytick={0,1,2,3,4,5},
        yticklabel style={font=\scriptsize},
        width=6cm,
    ]

    \addplot[color=CustomRed, mark=triangle*, line width=0.8pt] coordinates {(25,4.425)(50,4.423)(100,4.421)(200,4.416)};
    
    \end{axis}
    \end{tikzpicture}
    \end{minipage}

    \caption{Objective evaluation results with different sampling steps in \modelname. We evaluate on a subset of MTD with inpainting gap = 1000ms. Results indicate \modelname has almost identical quality when we use less sampling steps.}
    
    \label{fig: Inpainting sampling steps}
\end{figure}

\section{Conclusion and Future Work}

This paper presents \modelname, a novel audio restoration model for music bandwidth extension and inpainting at 44.1kHz. We present an end-to-end solution requiring no vocoder or codec. We demonstrate more stable learning dynamics with a factorized magnitude-phase STFT representation, while also guaranteeing proper phase values with SVD orthogonalization. We use a two-stage training strategy and a $t$-range partitioning method to improve quality, as well as a sampling method for very long audio restoration. We also curated a collection of permissively licensed high quality music data to train our model. Extensive experiments show that \modelname achieves the state-of-the-art quality on several OOD test sets, validating the effectiveness and generalization ability of our model.

For future works, we plan to: \textit{(1)} extend \modelname to support more audio restoration tasks beyond bandwidth extension and inpainting, \textit{(2)} investigate using the most recent advances in large-scale model architectures to scale-up our model for better quality, generalization, and emergent abilities, \textit{(3)} investigate better usage of training data in the setup where we combine data from very different data distributions, and \textit{(4)} extend our solution to stereo music.

\section*{Acknowledgments}

We thank Siddharth Gururani and Pin-Jui Ku for helpful discussions at the early stages of this work.

\newpage
\bibliography{main.bib}

\begin{thebibliography}{79}
\providecommand{\natexlab}[1]{#1}
\providecommand{\url}[1]{\texttt{#1}}
\expandafter\ifx\csname urlstyle\endcsname\relax
  \providecommand{\doi}[1]{doi: #1}\else
  \providecommand{\doi}{doi: \begingroup \urlstyle{rm}\Url}\fi

\bibitem[Agostinelli et~al.(2023)Agostinelli, Denk, Borsos, Engel, Verzetti,
  Caillon, Huang, Jansen, Roberts, Tagliasacchi,
  et~al.]{agostinelli2023musiclm}
Andrea Agostinelli, Timo~I Denk, Zal{\'a}n Borsos, Jesse Engel, Mauro Verzetti,
  Antoine Caillon, Qingqing Huang, Aren Jansen, Adam Roberts, Marco
  Tagliasacchi, et~al.
\newblock Musiclm: Generating music from text.
\newblock \emph{arXiv preprint arXiv:2301.11325}, 2023.

\bibitem[Albergo et~al.(2023)Albergo, Boffi, and
  Vanden-Eijnden]{albergo2023stochastic}
Michael~S Albergo, Nicholas~M Boffi, and Eric Vanden-Eijnden.
\newblock Stochastic interpolants: A unifying framework for flows and
  diffusions.
\newblock \emph{arXiv preprint arXiv:2303.08797}, 2023.

\bibitem[Anderson(1982)]{anderson1982reverse}
Brian~DO Anderson.
\newblock Reverse-time diffusion equation models.
\newblock \emph{Stochastic Processes and their Applications}, 12\penalty0
  (3):\penalty0 313--326, 1982.

\bibitem[Asaad et~al.(2024)Asaad, Jacquelin, Perrotin, Girin, and
  Hueber]{asaad2024fill}
Ihab Asaad, Maxime Jacquelin, Olivier Perrotin, Laurent Girin, and Thomas
  Hueber.
\newblock Fill in the gap! combining self-supervised representation learning
  with neural audio synthesis for speech inpainting.
\newblock \emph{arXiv preprint arXiv:2405.20101}, 2024.

\bibitem[Balaji et~al.(2022)Balaji, Nah, Huang, Vahdat, Song, Zhang, Kreis,
  Aittala, Aila, Laine, et~al.]{balaji2022ediff}
Yogesh Balaji, Seungjun Nah, Xun Huang, Arash Vahdat, Jiaming Song, Qinsheng
  Zhang, Karsten Kreis, Miika Aittala, Timo Aila, Samuli Laine, et~al.
\newblock ediff-i: Text-to-image diffusion models with an ensemble of expert
  denoisers.
\newblock \emph{arXiv preprint arXiv:2211.01324}, 2022.

\bibitem[Bar-Tal et~al.(2023)Bar-Tal, Yariv, Lipman, and
  Dekel]{bar2023multidiffusion}
Omer Bar-Tal, Lior Yariv, Yaron Lipman, and Tali Dekel.
\newblock Multidiffusion: Fusing diffusion paths for controlled image
  generation.
\newblock \emph{arXiv preprint arXiv:2302.08113}, 2023.

\bibitem[Bogdanov et~al.(2019)Bogdanov, Won, Tovstogan, Porter, and
  Serra]{bogdanov2019mtg}
Dmitry Bogdanov, Minz Won, Philip Tovstogan, Alastair Porter, and Xavier Serra.
\newblock The mtg-jamendo dataset for automatic music tagging.
\newblock In \emph{Machine Learning for Music Discovery Workshop, International
  Conference on Machine Learning (ICML 2019)}, Long Beach, CA, United States,
  2019.
\newblock URL \url{http://hdl.handle.net/10230/42015}.

\bibitem[Borsos et~al.(2022)Borsos, Sharifi, and
  Tagliasacchi]{borsos2022speechpainter}
Zal{\'a}n Borsos, Matt Sharifi, and Marco Tagliasacchi.
\newblock Speechpainter: Text-conditioned speech inpainting.
\newblock \emph{arXiv preprint arXiv:2202.07273}, 2022.

\bibitem[Chen \& Lipman(2023)Chen and Lipman]{chen2023riemannian}
Ricky~TQ Chen and Yaron Lipman.
\newblock Riemannian flow matching on general geometries.
\newblock \emph{arXiv preprint arXiv:2302.03660}, 2023.

\bibitem[Chen et~al.(2021)Chen, Liu, and Theodorou]{chen2021likelihood}
Tianrong Chen, Guan-Horng Liu, and Evangelos~A Theodorou.
\newblock Likelihood training of schr$\backslash$" odinger bridge using
  forward-backward sdes theory.
\newblock \emph{arXiv preprint arXiv:2110.11291}, 2021.

\bibitem[Chinen et~al.(2020)Chinen, Lim, Skoglund, Gureev, O'Gorman, and
  Hines]{chinen2020visqol}
Michael Chinen, Felicia~SC Lim, Jan Skoglund, Nikita Gureev, Feargus O'Gorman,
  and Andrew Hines.
\newblock Visqol v3: An open source production ready objective speech and audio
  metric.
\newblock In \emph{2020 twelfth international conference on quality of
  multimedia experience (QoMEX)}, pp.\  1--6. IEEE, 2020.

\bibitem[Chung et~al.(2022{\natexlab{a}})Chung, Kim, Mccann, Klasky, and
  Ye]{chung2022diffusion}
Hyungjin Chung, Jeongsol Kim, Michael~T Mccann, Marc~L Klasky, and Jong~Chul
  Ye.
\newblock Diffusion posterior sampling for general noisy inverse problems.
\newblock \emph{arXiv preprint arXiv:2209.14687}, 2022{\natexlab{a}}.

\bibitem[Chung et~al.(2022{\natexlab{b}})Chung, Sim, Ryu, and
  Ye]{chung2022improving}
Hyungjin Chung, Byeongsu Sim, Dohoon Ryu, and Jong~Chul Ye.
\newblock Improving diffusion models for inverse problems using manifold
  constraints.
\newblock \emph{Advances in Neural Information Processing Systems},
  35:\penalty0 25683--25696, 2022{\natexlab{b}}.

\bibitem[De~Bortoli et~al.(2021)De~Bortoli, Thornton, Heng, and
  Doucet]{de2021diffusion}
Valentin De~Bortoli, James Thornton, Jeremy Heng, and Arnaud Doucet.
\newblock Diffusion schr{\"o}dinger bridge with applications to score-based
  generative modeling.
\newblock \emph{Advances in Neural Information Processing Systems},
  34:\penalty0 17695--17709, 2021.

\bibitem[Defferrard et~al.(2016)Defferrard, Benzi, Vandergheynst, and
  Bresson]{defferrard2016fma}
Micha{\"e}l Defferrard, Kirell Benzi, Pierre Vandergheynst, and Xavier Bresson.
\newblock Fma: A dataset for music analysis.
\newblock \emph{arXiv preprint arXiv:1612.01840}, 2016.

\bibitem[Dhariwal \& Nichol(2021)Dhariwal and Nichol]{dhariwal2021diffusion}
Prafulla Dhariwal and Alexander Nichol.
\newblock Diffusion models beat gans on image synthesis.
\newblock \emph{Advances in neural information processing systems},
  34:\penalty0 8780--8794, 2021.

\bibitem[Engel et~al.(2017)Engel, Resnick, Roberts, Dieleman, Norouzi, Eck, and
  Simonyan]{engel2017neural}
Jesse Engel, Cinjon Resnick, Adam Roberts, Sander Dieleman, Mohammad Norouzi,
  Douglas Eck, and Karen Simonyan.
\newblock Neural audio synthesis of musical notes with wavenet autoencoders.
\newblock In \emph{International Conference on Machine Learning}, pp.\
  1068--1077. PMLR, 2017.

\bibitem[Erell \& Weintraub(1990)Erell and Weintraub]{erell1990estimation}
Adoram Erell and Mitch Weintraub.
\newblock Estimation using log-spectral-distance criterion for noise-robust
  speech recognition.
\newblock In \emph{International Conference on Acoustics, Speech, and Signal
  Processing}, pp.\  853--856. IEEE, 1990.

\bibitem[Hawthorne et~al.(2019)Hawthorne, Stasyuk, Roberts, Simon, Huang,
  Dieleman, Elsen, Engel, and Eck]{hawthorne2018enabling}
Curtis Hawthorne, Andriy Stasyuk, Adam Roberts, Ian Simon, Cheng-Zhi~Anna
  Huang, Sander Dieleman, Erich Elsen, Jesse Engel, and Douglas Eck.
\newblock Enabling factorized piano music modeling and generation with the
  {MAESTRO} dataset.
\newblock In \emph{International Conference on Learning Representations}, 2019.
\newblock URL \url{https://openreview.net/forum?id=r1lYRjC9F7}.

\bibitem[Ho et~al.(2020)Ho, Jain, and Abbeel]{ho2020denoising}
Jonathan Ho, Ajay Jain, and Pieter Abbeel.
\newblock Denoising diffusion probabilistic models.
\newblock \emph{Advances in neural information processing systems},
  33:\penalty0 6840--6851, 2020.

\bibitem[Ho et~al.(2022)Ho, Salimans, Gritsenko, Chan, Norouzi, and
  Fleet]{ho2022video}
Jonathan Ho, Tim Salimans, Alexey Gritsenko, William Chan, Mohammad Norouzi,
  and David~J Fleet.
\newblock Video diffusion models.
\newblock \emph{Advances in Neural Information Processing Systems},
  35:\penalty0 8633--8646, 2022.

\bibitem[Juki{\'c} et~al.(2024)Juki{\'c}, Korostik, Balam, and
  Ginsburg]{jukic2024schr}
Ante Juki{\'c}, Roman Korostik, Jagadeesh Balam, and Boris Ginsburg.
\newblock Schrödinger bridge for generative speech enhancement.
\newblock \emph{arXiv preprint arXiv:2407.16074}, 2024.

\bibitem[Kim et~al.(2024)Kim, Lee, Choi, and Lee]{kim2024audio}
Seung-Bin Kim, Sang-Hoon Lee, Ha-Yeong Choi, and Seong-Whan Lee.
\newblock Audio super-resolution with robust speech representation learning of
  masked autoencoder.
\newblock \emph{IEEE/ACM Transactions on Audio, Speech, and Language
  Processing}, 2024.

\bibitem[Kong et~al.(2021)Kong, Ping, Huang, Zhao, and Catanzaro]{kongdiffwave}
Zhifeng Kong, Wei Ping, Jiaji Huang, Kexin Zhao, and Bryan Catanzaro.
\newblock Diffwave: A versatile diffusion model for audio synthesis.
\newblock In \emph{International Conference on Learning Representations}, 2021.

\bibitem[Ku et~al.(2024)Ku, Liu, Korostik, Huang, Fu, and Jukić]{ku2024gen}
Pin-Jui Ku, Alexander~H. Liu, Roman Korostik, Sung-Feng Huang, Szu-Wei Fu, and
  Ante Jukić.
\newblock Generative speech foundation model pretraining for high-quality
  speech extraction and restoration, 2024.
\newblock URL \url{https://arxiv.org/abs/2409.16117}.

\bibitem[Lee \& Han(2021)Lee and Han]{lee2021nu}
Junhyeok Lee and Seungu Han.
\newblock Nu-wave: A diffusion probabilistic model for neural audio upsampling.
\newblock \emph{arXiv preprint arXiv:2104.02321}, 2021.

\bibitem[Lee et~al.(2023)Lee, Ping, Ginsburg, Catanzaro, and
  Yoon]{lee2023bigvgan}
Sang-gil Lee, Wei Ping, Boris Ginsburg, Bryan Catanzaro, and Sungroh Yoon.
\newblock Bigvgan: A universal neural vocoder with large-scale training.
\newblock In \emph{The Eleventh International Conference on Learning
  Representations}, 2023.
\newblock URL \url{https://openreview.net/forum?id=iTtGCMDEzS_}.

\bibitem[Lee et~al.(2024)Lee, Kong, Goel, Kim, Valle, and
  Catanzaro]{lee2024etta}
Sang-gil Lee, Zhifeng Kong, Arushi Goel, Sungwon Kim, Rafael Valle, and Bryan
  Catanzaro.
\newblock Etta: Elucidating the design space of text-to-audio models.
\newblock \emph{arXiv preprint arXiv:2412.19351}, 2024.

\bibitem[Lemercier et~al.(2024)Lemercier, Richter, Welker, Moliner,
  V{\"a}lim{\"a}ki, and Gerkmann]{lemercier2024diffusion}
Jean-Marie Lemercier, Julius Richter, Simon Welker, Eloi Moliner, Vesa
  V{\"a}lim{\"a}ki, and Timo Gerkmann.
\newblock Diffusion models for audio restoration.
\newblock \emph{arXiv preprint arXiv:2402.09821}, 2024.

\bibitem[L{\'e}onard(2013)]{leonard2013survey}
Christian L{\'e}onard.
\newblock A survey of the schr\"odinger problem and some of its connections
  with optimal transport.
\newblock \emph{arXiv preprint arXiv:1308.0215}, 2013.

\bibitem[Levinson et~al.(2020)Levinson, Esteves, Chen, Snavely, Kanazawa,
  Rostamizadeh, and Makadia]{levinsonSVD}
Jake Levinson, Carlos Esteves, Kefan Chen, Noah Snavely, Angjoo Kanazawa,
  Afshin Rostamizadeh, and Ameesh Makadia.
\newblock An analysis of svd for deep rotation estimation.
\newblock In H.~Larochelle, M.~Ranzato, R.~Hadsell, M.F. Balcan, and H.~Lin
  (eds.), \emph{Advances in Neural Information Processing Systems}, volume~33,
  pp.\  22554--22565. Curran Associates, Inc., 2020.

\bibitem[Li et~al.(2025)Li, Chen, Bao, and Zhu]{li2025bridge}
Chang Li, Zehua Chen, Fan Bao, and Jun Zhu.
\newblock Bridge-sr: Schr$\backslash$" odinger bridge for efficient sr.
\newblock \emph{arXiv preprint arXiv:2501.07897}, 2025.

\bibitem[Lipman et~al.(2022)Lipman, Chen, Ben-Hamu, Nickel, and
  Le]{lipman2022flow}
Yaron Lipman, Ricky~TQ Chen, Heli Ben-Hamu, Maximilian Nickel, and Matt Le.
\newblock Flow matching for generative modeling.
\newblock \emph{arXiv preprint arXiv:2210.02747}, 2022.

\bibitem[Liu et~al.(2023{\natexlab{a}})Liu, Vahdat, Huang, Theodorou, Nie, and
  Anandkumar]{liuI2SB}
Guan-Horng Liu, Arash Vahdat, De-An Huang, Evangelos~A. Theodorou, Weili Nie,
  and Anima Anandkumar.
\newblock I2sb: image-to-image schrödinger bridge.
\newblock In \emph{Proceedings of the 40th International Conference on Machine
  Learning}, ICML'23. JMLR.org, 2023{\natexlab{a}}.

\bibitem[Liu et~al.(2021)Liu, Kong, Tian, Zhao, Wang, Huang, and
  Wang]{liu2021voicefixer}
Haohe Liu, Qiuqiang Kong, Qiao Tian, Yan Zhao, DeLiang Wang, Chuanzeng Huang,
  and Yuxuan Wang.
\newblock Voicefixer: Toward general speech restoration with neural vocoder.
\newblock \emph{arXiv preprint arXiv:2109.13731}, 2021.

\bibitem[Liu et~al.(2022)Liu, Choi, Liu, Kong, Tian, and Wang]{liu2022neural}
Haohe Liu, Woosung Choi, Xubo Liu, Qiuqiang Kong, Qiao Tian, and DeLiang Wang.
\newblock Neural vocoder is all you need for speech super-resolution.
\newblock \emph{arXiv preprint arXiv:2203.14941}, 2022.

\bibitem[Liu et~al.(2024)Liu, Chen, Tian, Wang, and Plumbley]{liu2024audiosr}
Haohe Liu, Ke~Chen, Qiao Tian, Wenwu Wang, and Mark~D Plumbley.
\newblock Audiosr: Versatile audio super-resolution at scale.
\newblock In \emph{ICASSP 2024-2024 IEEE International Conference on Acoustics,
  Speech and Signal Processing (ICASSP)}, pp.\  1076--1080. IEEE, 2024.

\bibitem[Liu et~al.(2023{\natexlab{b}})Liu, Gan, and Yuan]{liu2023maid}
Kaiyang Liu, Wendong Gan, and Chenchen Yuan.
\newblock Maid: A conditional diffusion model for long music audio inpainting.
\newblock In \emph{ICASSP 2023-2023 IEEE International Conference on Acoustics,
  Speech and Signal Processing (ICASSP)}, pp.\  1--5. IEEE, 2023{\natexlab{b}}.

\bibitem[Liutkus et~al.(2014)Liutkus, Fitzgerald, Rafii, Pardo, and
  Daudet]{liutkus2014kernel}
Antoine Liutkus, Derry Fitzgerald, Zafar Rafii, Bryan Pardo, and Laurent
  Daudet.
\newblock Kernel additive models for source separation.
\newblock \emph{IEEE Transactions on Signal Processing}, 62\penalty0
  (16):\penalty0 4298--4310, 2014.

\bibitem[Lostanlen \& Cella(2016)Lostanlen and Cella]{lostanlen2016deep}
Vincent Lostanlen and Carmine-Emanuele Cella.
\newblock Deep convolutional networks on the pitch spiral for musical
  instrument recognition.
\newblock \emph{arXiv preprint arXiv:1605.06644}, 2016.

\bibitem[Manilow et~al.(2019)Manilow, Wichern, Seetharaman, and
  Le~Roux]{manilow2019cutting}
Ethan Manilow, Gordon Wichern, Prem Seetharaman, and Jonathan Le~Roux.
\newblock Cutting music source separation some {Slakh}: A dataset to study the
  impact of training data quality and quantity.
\newblock In \emph{Proc. IEEE Workshop on Applications of Signal Processing to
  Audio and Acoustics (WASPAA)}. IEEE, 2019.

\bibitem[Marafioti et~al.(2019)Marafioti, Perraudin, Holighaus, and
  Majdak]{marafioti2019context}
Andr{\'e}s Marafioti, Nathana{\"e}l Perraudin, Nicki Holighaus, and Piotr
  Majdak.
\newblock A context encoder for audio inpainting.
\newblock \emph{IEEE/ACM Transactions on Audio, Speech, and Language
  Processing}, 27\penalty0 (12):\penalty0 2362--2372, 2019.

\bibitem[Marafioti et~al.(2020)Marafioti, Majdak, Holighaus, and
  Perraudin]{marafioti2020gacela}
Andr{\'e}s Marafioti, Piotr Majdak, Nicki Holighaus, and Nathana{\"e}l
  Perraudin.
\newblock Gacela: A generative adversarial context encoder for long audio
  inpainting of music.
\newblock \emph{IEEE Journal of Selected Topics in Signal Processing},
  15\penalty0 (1):\penalty0 120--131, 2020.

\bibitem[Moliner \& V{\"a}lim{\"a}ki(2022)Moliner and
  V{\"a}lim{\"a}ki]{moliner2022behm}
Eloi Moliner and Vesa V{\"a}lim{\"a}ki.
\newblock Behm-gan: Bandwidth extension of historical music using generative
  adversarial networks.
\newblock \emph{IEEE/ACM Transactions on Audio, Speech, and Language
  Processing}, 31:\penalty0 943--956, 2022.

\bibitem[Moliner \& V{\"a}lim{\"a}ki(2023)Moliner and
  V{\"a}lim{\"a}ki]{moliner2023diffusion}
Eloi Moliner and Vesa V{\"a}lim{\"a}ki.
\newblock Diffusion-based audio inpainting.
\newblock \emph{arXiv preprint arXiv:2305.15266}, 2023.

\bibitem[Moliner et~al.(2023)Moliner, Lehtinen, and
  V{\"a}lim{\"a}ki]{moliner2023solving}
Eloi Moliner, Jaakko Lehtinen, and Vesa V{\"a}lim{\"a}ki.
\newblock Solving audio inverse problems with a diffusion model.
\newblock In \emph{ICASSP 2023-2023 IEEE International Conference on Acoustics,
  Speech and Signal Processing (ICASSP)}, pp.\  1--5. IEEE, 2023.

\bibitem[Mutlu(2024)]{musicinstrument}
Abdulvahap Mutlu.
\newblock Music instrument sounds for classification.
\newblock \emph{Kaggle}, 2024.

\bibitem[Narayanaswamy et~al.(2021)Narayanaswamy, Thiagarajan, and
  Spanias]{narayanaswamy2021design}
Vivek~Sivaraman Narayanaswamy, Jayaraman~J Thiagarajan, and Andreas Spanias.
\newblock On the design of deep priors for unsupervised audio restoration.
\newblock \emph{arXiv preprint arXiv:2104.07161}, 2021.

\bibitem[Ostermann et~al.(2023)Ostermann, Vatolkin, and
  Ebeling]{ostermann2023aam}
Fabian Ostermann, Igor Vatolkin, and Martin Ebeling.
\newblock Aam: a dataset of artificial audio multitracks for diverse music
  information retrieval tasks.
\newblock \emph{EURASIP Journal on Audio, Speech, and Music Processing},
  2023\penalty0 (1):\penalty0 13, 2023.

\bibitem[Ouyang et~al.(2022)Ouyang, Wu, Jiang, Almeida, Wainwright, Mishkin,
  Zhang, Agarwal, Slama, Ray, et~al.]{ouyang2022training}
Long Ouyang, Jeffrey Wu, Xu~Jiang, Diogo Almeida, Carroll Wainwright, Pamela
  Mishkin, Chong Zhang, Sandhini Agarwal, Katarina Slama, Alex Ray, et~al.
\newblock Training language models to follow instructions with human feedback.
\newblock \emph{Advances in neural information processing systems},
  35:\penalty0 27730--27744, 2022.

\bibitem[Peebles \& Xie(2023)Peebles and Xie]{peebles2023scalable}
William Peebles and Saining Xie.
\newblock Scalable diffusion models with transformers.
\newblock In \emph{Proceedings of the IEEE/CVF International Conference on
  Computer Vision}, pp.\  4195--4205, 2023.

\bibitem[Peer \& Gerkmann(2022)Peer and Gerkmann]{peer2022phase}
Tal Peer and Timo Gerkmann.
\newblock Phase-aware deep speech enhancement: It's all about the frame length.
\newblock \emph{JASA Express Letters}, 2\penalty0 (10), 2022.

\bibitem[Polyak et~al.(2024)Polyak, Zohar, Brown, Tjandra, Sinha, Lee, Vyas,
  Shi, Ma, Chuang, et~al.]{polyak2024movie}
Adam Polyak, Amit Zohar, Andrew Brown, Andros Tjandra, Animesh Sinha, Ann Lee,
  Apoorv Vyas, Bowen Shi, Chih-Yao Ma, Ching-Yao Chuang, et~al.
\newblock Movie gen: A cast of media foundation models.
\newblock \emph{arXiv preprint arXiv:2410.13720}, 2024.

\bibitem[Raffel et~al.()Raffel, McFee, Humphrey, Salamon, Nieto, Liang, and
  Ellis]{raffelmir_eval}
Colin Raffel, Brian McFee, Eric~J Humphrey, Justin Salamon, Oriol Nieto, Dawen
  Liang, and Daniel~PW Ellis.
\newblock mir\_eval.

\bibitem[Richter et~al.(2023)Richter, Welker, Lemercier, Lay, and
  Gerkmann]{richter2023sgmse}
Julius Richter, Simon Welker, Jean-Marie Lemercier, Bunlong Lay, and Timo
  Gerkmann.
\newblock Speech enhancement and dereverberation with diffusion-based
  generative models.
\newblock \emph{IEEE/ACM Trans. on Audio, Speech, and Language Process.},
  31:\penalty0 2351--2364, 2023.

\bibitem[Roberts(2022)]{pianotriads}
David Roberts.
\newblock Piano triads wavset.
\newblock \emph{Kaggle}, 2022.

\bibitem[Rombach et~al.(2022)Rombach, Blattmann, Lorenz, Esser, and
  Ommer]{rombach2022high}
Robin Rombach, Andreas Blattmann, Dominik Lorenz, Patrick Esser, and Bj{\"o}rn
  Ommer.
\newblock High-resolution image synthesis with latent diffusion models.
\newblock In \emph{Proceedings of the IEEE/CVF conference on computer vision
  and pattern recognition}, pp.\  10684--10695, 2022.

\bibitem[Ronneberger et~al.(2015)Ronneberger, Fischer, and
  Brox]{ronneberger2015u}
Olaf Ronneberger, Philipp Fischer, and Thomas Brox.
\newblock U-net: Convolutional networks for biomedical image segmentation.
\newblock In \emph{Medical image computing and computer-assisted
  intervention--MICCAI 2015: 18th international conference, Munich, Germany,
  October 5-9, 2015, proceedings, part III 18}, pp.\  234--241. Springer, 2015.

\bibitem[Saharia et~al.(2022)Saharia, Ho, Chan, Salimans, Fleet, and
  Norouzi]{saharia2022image}
Chitwan Saharia, Jonathan Ho, William Chan, Tim Salimans, David~J Fleet, and
  Mohammad Norouzi.
\newblock Image super-resolution via iterative refinement.
\newblock \emph{IEEE transactions on pattern analysis and machine
  intelligence}, 45\penalty0 (4):\penalty0 4713--4726, 2022.

\bibitem[Sch{\"o}nemann(1966)]{schonemann1966generalized}
Peter~H Sch{\"o}nemann.
\newblock A generalized solution of the orthogonal procrustes problem.
\newblock \emph{Psychometrika}, 31\penalty0 (1):\penalty0 1--10, 1966.

\bibitem[Schr{\"o}dinger(1932)]{schrodinger1932theorie}
Erwin Schr{\"o}dinger.
\newblock Sur la th{\'e}orie relativiste de l'{\'e}lectron et
  l'interpr{\'e}tation de la m{\'e}canique quantique.
\newblock In \emph{Annales de l'institut Henri Poincar{\'e}}, volume~2, pp.\
  269--310, 1932.

\bibitem[Serr{\`a} et~al.(2022)Serr{\`a}, Pascual, Pons, Araz, and
  Scaini]{serra2022universal}
Joan Serr{\`a}, Santiago Pascual, Jordi Pons, R~Oguz Araz, and Davide Scaini.
\newblock Universal speech enhancement with score-based diffusion.
\newblock \emph{arXiv preprint arXiv:2206.03065}, 2022.

\bibitem[Shuai et~al.(2023)Shuai, Shi, Gan, and Liu]{shuai2023mdctgan}
Chenhao Shuai, Chaohua Shi, Lu~Gan, and Hongqing Liu.
\newblock mdctgan: Taming transformer-based gan for speech super-resolution
  with modified dct spectra.
\newblock \emph{arXiv preprint arXiv:2305.11104}, 2023.

\bibitem[Snyder et~al.(2015)Snyder, Chen, and Povey]{snyder2015musan}
David Snyder, Guoguo Chen, and Daniel Povey.
\newblock Musan: A music, speech, and noise corpus.
\newblock \emph{arXiv preprint arXiv:1510.08484}, 2015.

\bibitem[Sohl-Dickstein et~al.(2015)Sohl-Dickstein, Weiss, Maheswaranathan, and
  Ganguli]{sohl2015deep}
Jascha Sohl-Dickstein, Eric Weiss, Niru Maheswaranathan, and Surya Ganguli.
\newblock Deep unsupervised learning using nonequilibrium thermodynamics.
\newblock In \emph{International conference on machine learning}, pp.\
  2256--2265. PMLR, 2015.

\bibitem[Song et~al.(2023)Song, Vahdat, Mardani, and
  Kautz]{song2023pseudoinverse}
Jiaming Song, Arash Vahdat, Morteza Mardani, and Jan Kautz.
\newblock Pseudoinverse-guided diffusion models for inverse problems.
\newblock In \emph{International Conference on Learning Representations}, 2023.

\bibitem[Song et~al.(2021)Song, Sohl-Dickstein, Kingma, Kumar, Ermon, and
  Poole]{songscore}
Yang Song, Jascha Sohl-Dickstein, Diederik~P Kingma, Abhishek Kumar, Stefano
  Ermon, and Ben Poole.
\newblock Score-based generative modeling through stochastic differential
  equations.
\newblock In \emph{International Conference on Learning Representations}, 2021.

\bibitem[Sturm(2013)]{sturm2013gtzan}
Bob~L Sturm.
\newblock The gtzan dataset: Its contents, its faults, their effects on
  evaluation, and its future use.
\newblock \emph{arXiv preprint arXiv:1306.1461}, 2013.

\bibitem[Su et~al.(2024)Su, Ahmed, Lu, Pan, Bo, and Liu]{su2024roformer}
Jianlin Su, Murtadha Ahmed, Yu~Lu, Shengfeng Pan, Wen Bo, and Yunfeng Liu.
\newblock Roformer: Enhanced transformer with rotary position embedding.
\newblock \emph{Neurocomputing}, 568:\penalty0 127063, 2024.

\bibitem[Thickstun et~al.(2017)Thickstun, Harchaoui, and
  Kakade]{thickstun2017learning}
John Thickstun, Zaid Harchaoui, and Sham~M. Kakade.
\newblock Learning features of music from scratch.
\newblock In \emph{International Conference on Learning Representations
  (ICLR)}, 2017.

\bibitem[Tong et~al.(2023)Tong, Malkin, Huguet, Zhang, Rector-Brooks, Fatras,
  Wolf, and Bengio]{tong2023conditional}
Alexander Tong, Nikolay Malkin, Guillaume Huguet, Yanlei Zhang, Jarrid
  Rector-Brooks, Kilian Fatras, Guy Wolf, and Yoshua Bengio.
\newblock Conditional flow matching: Simulation-free dynamic optimal transport.
\newblock \emph{arXiv preprint arXiv:2302.00482}, 2023.

\bibitem[Wang et~al.(2024)Wang, Liu, Harper, Kendrick, Salzmann, and
  Cernak]{wang2024diffusion}
Siyi Wang, Siyi Liu, Andrew Harper, Paul Kendrick, Mathieu Salzmann, and Milos
  Cernak.
\newblock Diffusion-based speech enhancement with schr$\backslash$" odinger
  bridge and symmetric noise schedule.
\newblock \emph{arXiv preprint arXiv:2409.05116}, 2024.

\bibitem[Wang et~al.(2023)Wang, Ju, Tan, He, Wu, Bian, et~al.]{wang2023audit}
Yuancheng Wang, Zeqian Ju, Xu~Tan, Lei He, Zhizheng Wu, Jiang Bian, et~al.
\newblock Audit: Audio editing by following instructions with latent diffusion
  models.
\newblock \emph{Advances in Neural Information Processing Systems},
  36:\penalty0 71340--71357, 2023.

\bibitem[Wu et~al.(2024)Wu, Marković, Krenn, Gebru, and Richard]{scoredec}
Yi-Chiao Wu, Dejan Marković, Steven Krenn, Israel~D. Gebru, and Alexander
  Richard.
\newblock Scoredec: A phase-preserving high-fidelity audio codec with a
  generalized score-based diffusion post-filter.
\newblock In \emph{ICASSP 2024 - 2024 IEEE International Conference on
  Acoustics, Speech and Signal Processing (ICASSP)}, pp.\  361--365, 2024.
\newblock \doi{10.1109/ICASSP48485.2024.10448371}.

\bibitem[Wu et~al.(2023)Wu, Chen, Zhang, Hui, Berg-Kirkpatrick, and
  Dubnov]{wu2023large}
Yusong Wu, Ke~Chen, Tianyu Zhang, Yuchen Hui, Taylor Berg-Kirkpatrick, and
  Shlomo Dubnov.
\newblock Large-scale contrastive language-audio pretraining with feature
  fusion and keyword-to-caption augmentation.
\newblock In \emph{ICASSP 2023-2023 IEEE International Conference on Acoustics,
  Speech and Signal Processing (ICASSP)}, pp.\  1--5. IEEE, 2023.

\bibitem[Xue et~al.(2022)Xue, Barua, Constant, Al-Rfou, Narang, Kale, Roberts,
  and Raffel]{xue2022byt5}
Linting Xue, Aditya Barua, Noah Constant, Rami Al-Rfou, Sharan Narang, Mihir
  Kale, Adam Roberts, and Colin Raffel.
\newblock Byt5: Towards a token-free future with pre-trained byte-to-byte
  models.
\newblock \emph{Transactions of the Association for Computational Linguistics},
  10:\penalty0 291--306, 2022.

\bibitem[Yu et~al.(2023)Yu, Yeh, Fazekas, and Tang]{yu2023conditioning}
Chin-Yun Yu, Sung-Lin Yeh, Gy{\"o}rgy Fazekas, and Hao Tang.
\newblock Conditioning and sampling in variational diffusion models for speech
  super-resolution.
\newblock In \emph{ICASSP 2023-2023 IEEE International Conference on Acoustics,
  Speech and Signal Processing (ICASSP)}, pp.\  1--5. IEEE, 2023.

\bibitem[Yun et~al.(2025)Yun, Kim, and Lee]{yun2025flowhigh}
Jun-Hak Yun, Seung-Bin Kim, and Seong-Whan Lee.
\newblock Flowhigh: Towards efficient and high-quality audio super-resolution
  with single-step flow matching.
\newblock \emph{arXiv preprint arXiv:2501.04926}, 2025.

\bibitem[Zalkow et~al.(2020)Zalkow, Balke, Arifi-M{\"u}ller, and
  M{\"u}ller]{zalkow2020mtd}
Frank Zalkow, Stefan Balke, Vlora Arifi-M{\"u}ller, and Meinard M{\"u}ller.
\newblock Mtd: A multimodal dataset of musical themes for mir research.
\newblock \emph{Trans. Int. Soc. Music. Inf. Retr.}, 3\penalty0 (1):\penalty0
  180--192, 2020.

\end{thebibliography}
\bibliographystyle{iclr2025_conference}

\newpage
\appendix

\section{More Samples on Bandwidth Extension}
\label{app: BWE samples}

\begin{figure}[!h]
    \centering
    \includegraphics[width=0.8\linewidth]{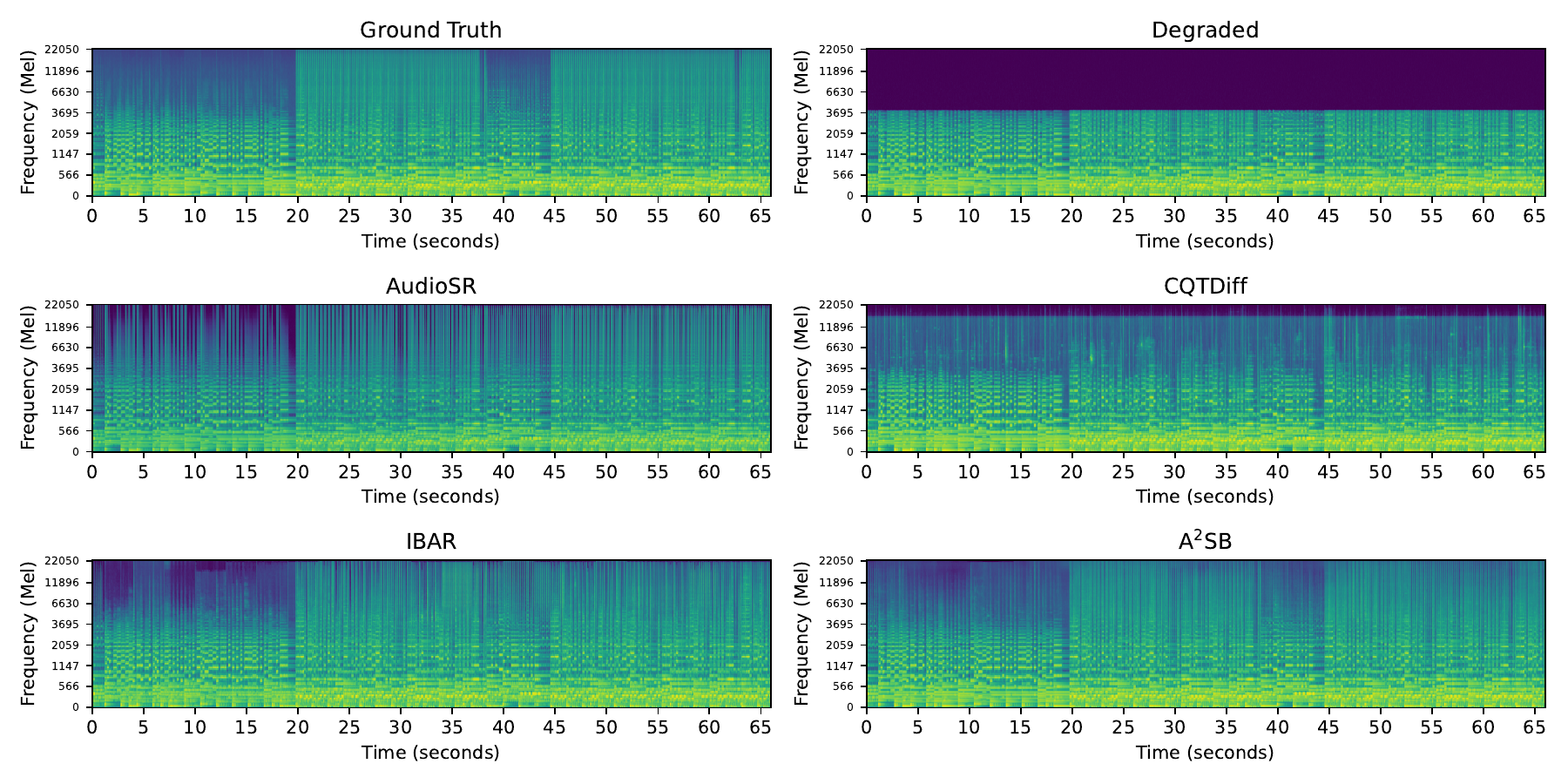}
    \caption{Qualitative comparison between different bandwidth extension methods with cutoff = 4kHz.}
    \label{fig: app: bwe 1}
\end{figure}

\begin{figure}[!h]
    \centering
    \includegraphics[width=0.8\linewidth]{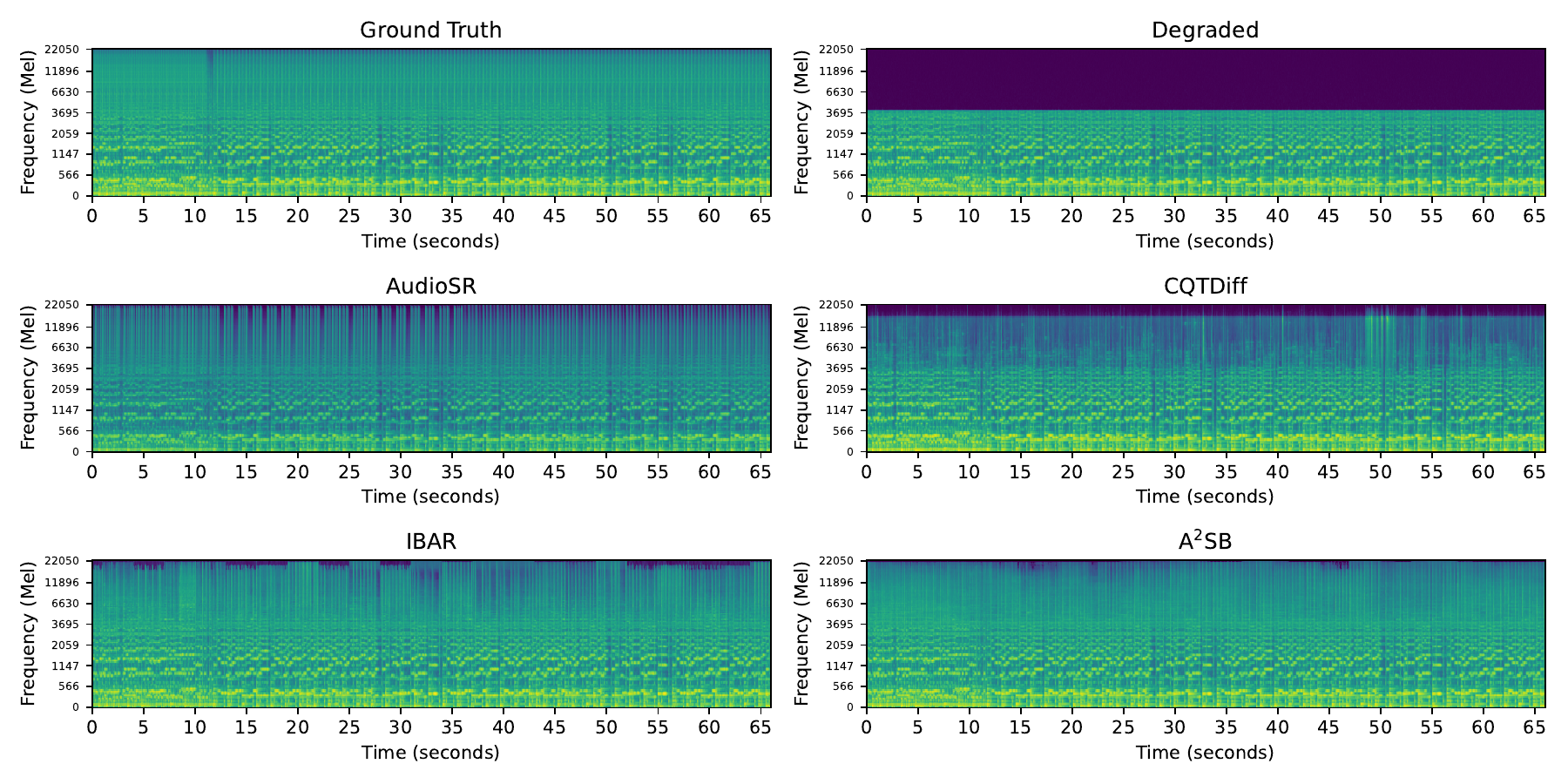}
    \caption{Qualitative comparison between different bandwidth extension methods with cutoff = 4kHz.}
    \label{fig: app: bwe 2}
\end{figure}

\begin{figure}[!h]
    \centering
    \includegraphics[width=0.8\linewidth]{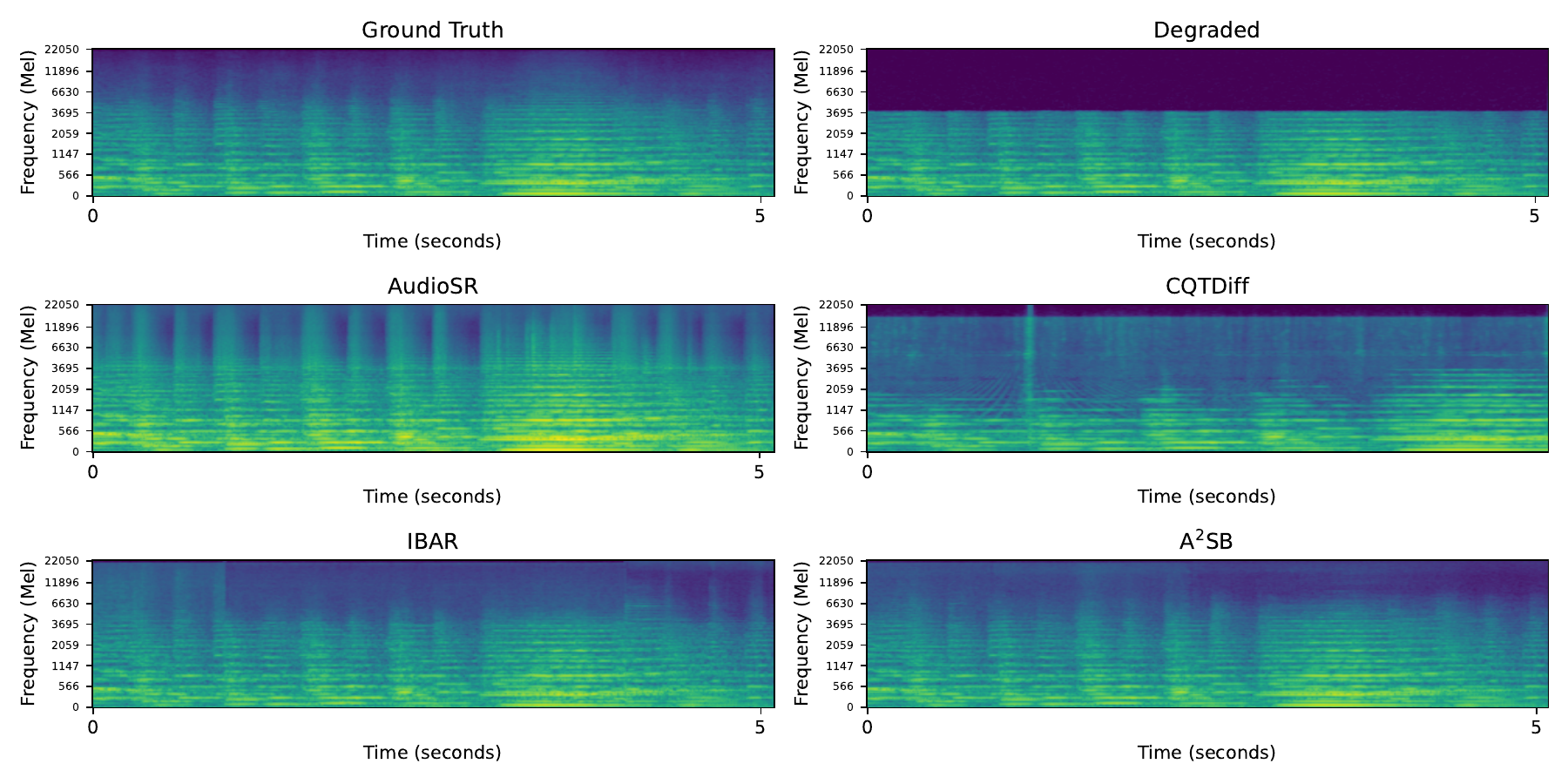}
    \caption{Qualitative comparison between different bandwidth extension methods with cutoff = 4kHz.}
    \label{fig: app: bwe 3}
\end{figure}

\newpage

\begin{figure}[!h]
    \centering
    \includegraphics[width=0.8\linewidth]{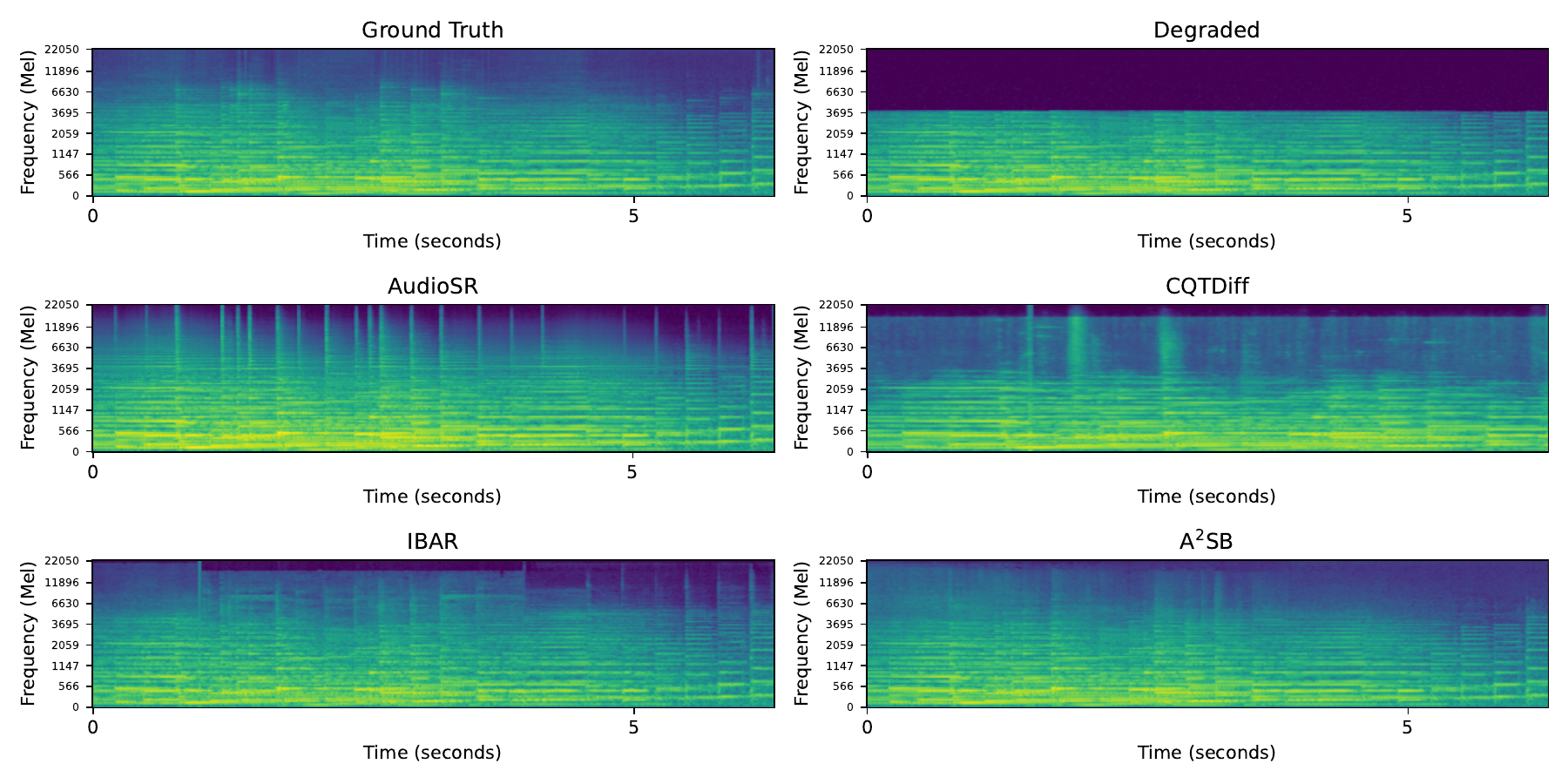}
    \caption{Qualitative comparison between different bandwidth extension methods with cutoff = 4kHz.}
    \label{fig: app: bwe 4}
\end{figure}

\begin{figure}[!h]
    \centering
    \includegraphics[width=0.8\linewidth]{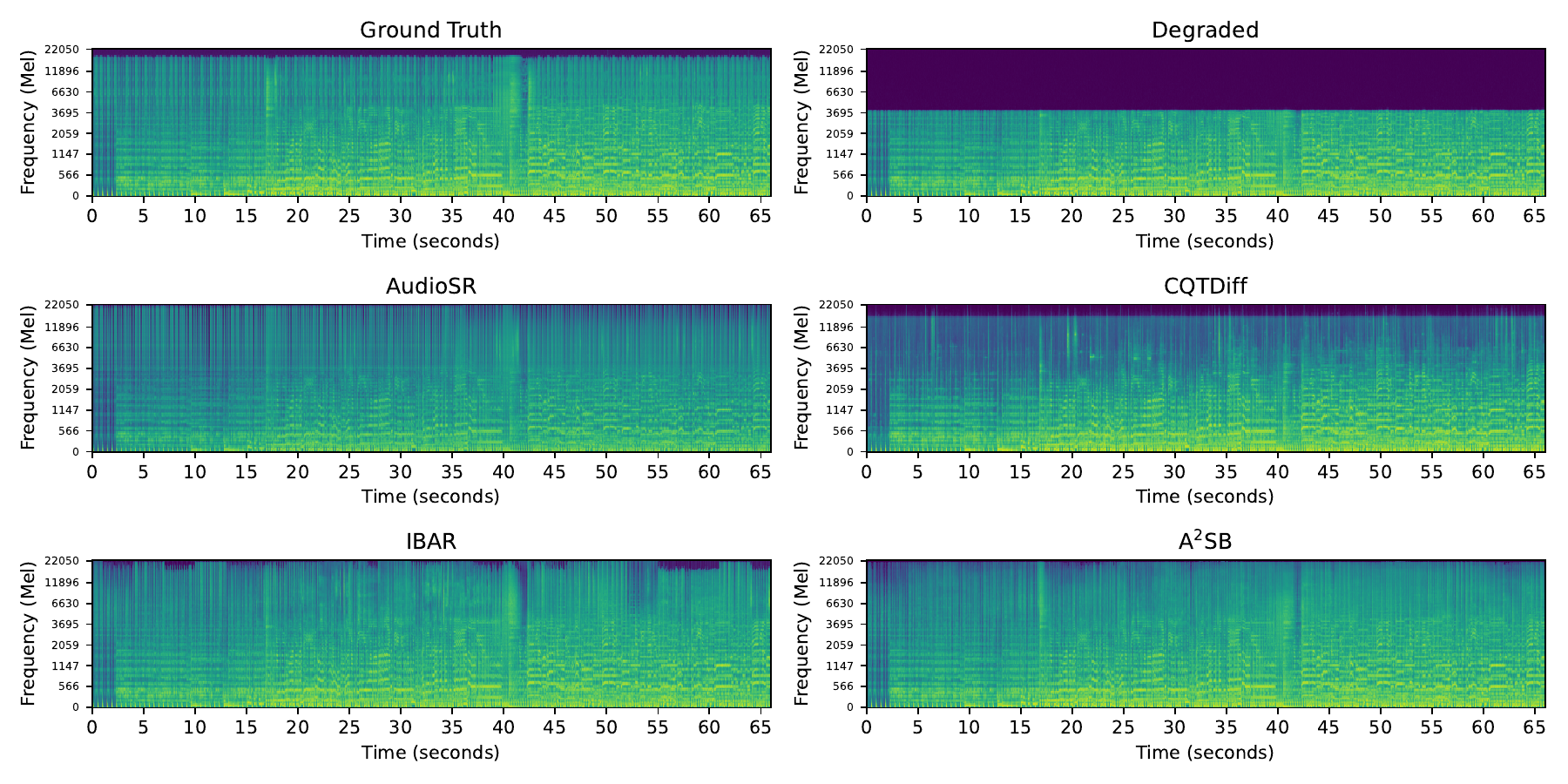}
    \caption{Qualitative comparison between different bandwidth extension methods with cutoff = 4kHz.}
    \label{fig: app: bwe 5}
\end{figure}

\begin{figure}[!h]
    \centering
    \includegraphics[width=0.8\linewidth]{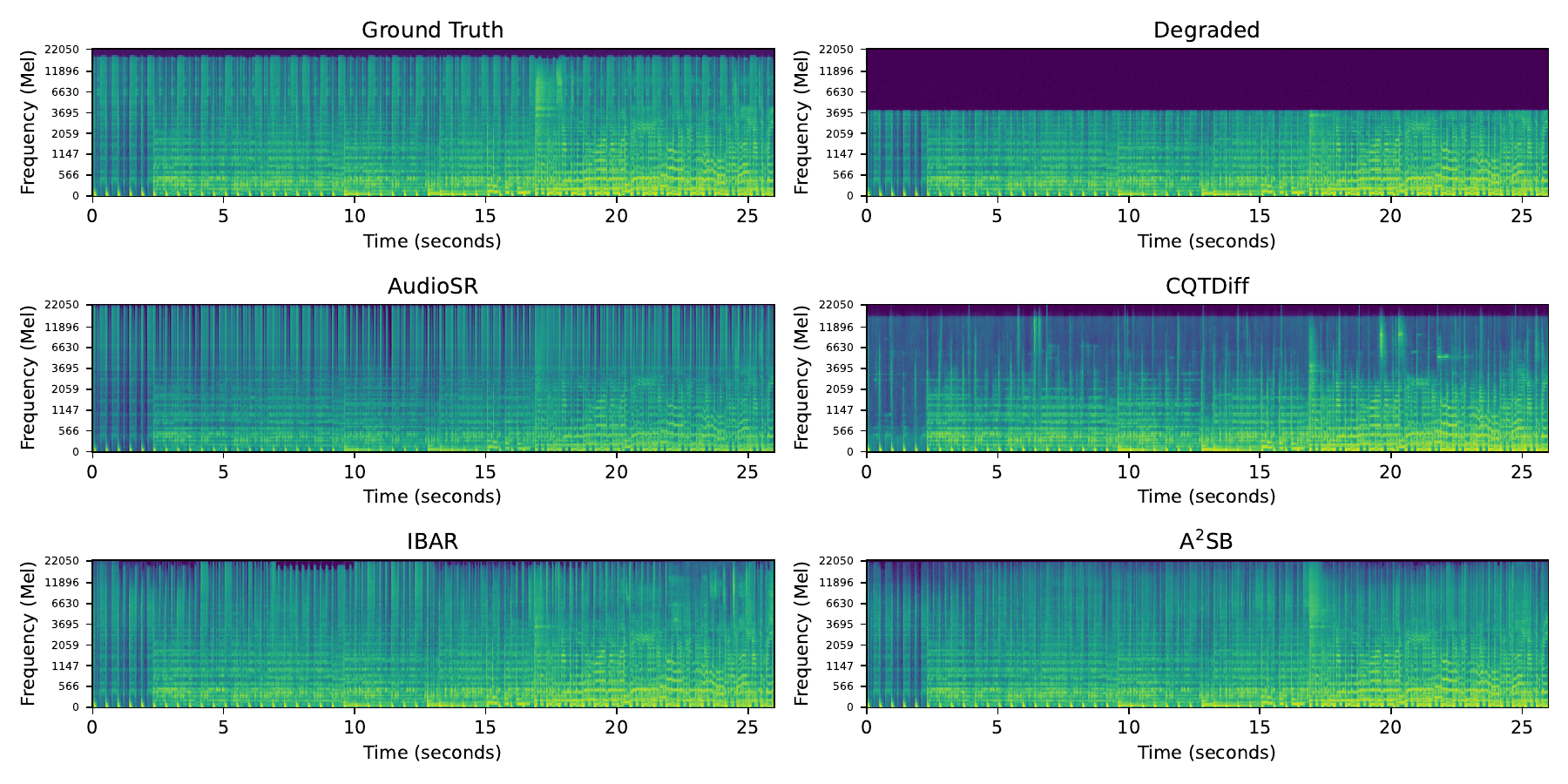}
    \caption{Qualitative comparison between different bandwidth extension methods with cutoff = 4kHz.}
    \label{fig: app: bwe 6}
\end{figure}

\newpage
\section{More Samples on Inpainting}
\label{app: INP samples}

\begin{figure}[!h]
    \centering
    \includegraphics[width=0.8\linewidth]{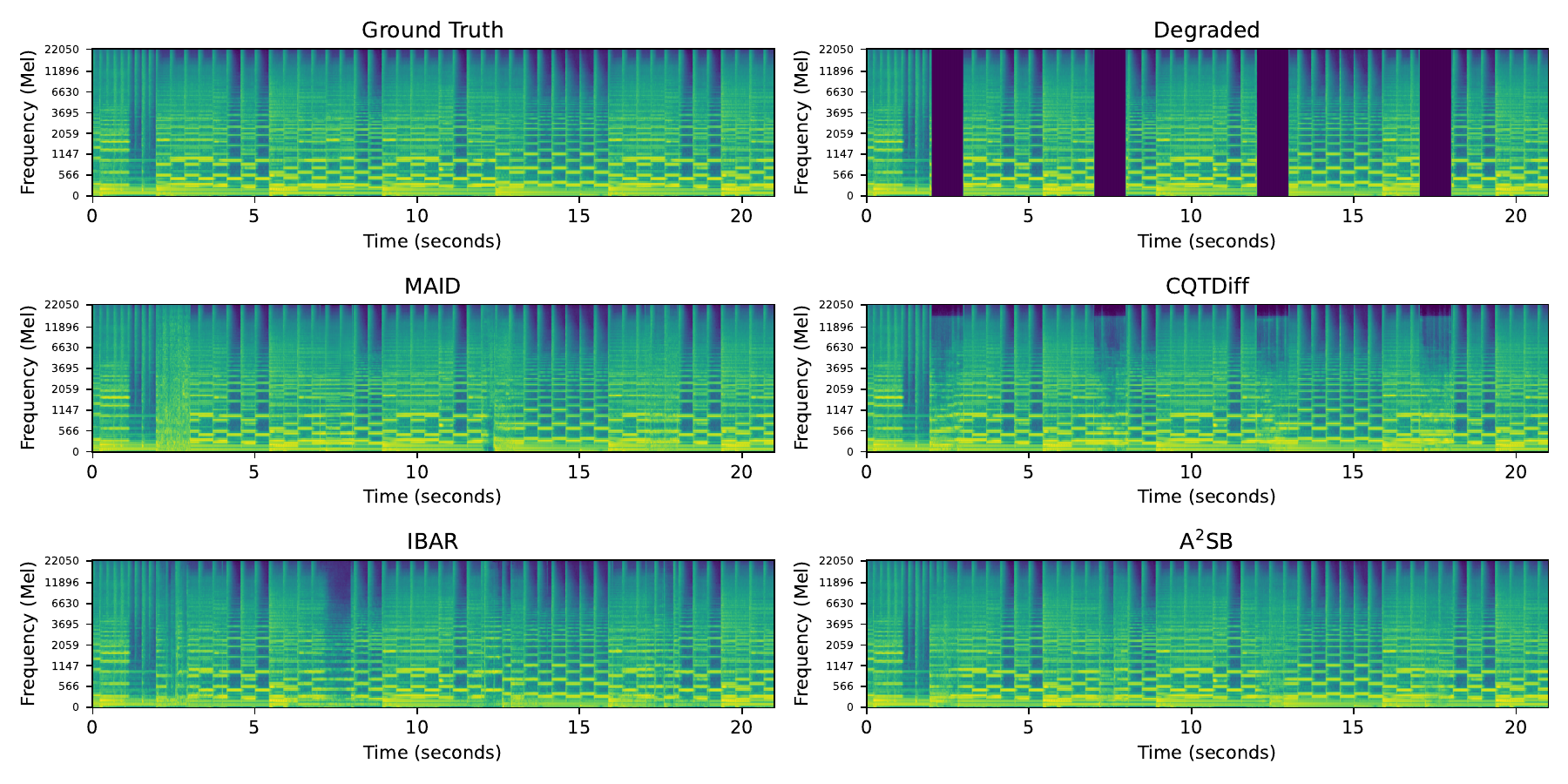}
    \caption{Qualitative comparison between different inpainting methods with inpainting gap = 1 sec.}
    \label{fig: app: inp 1}
\end{figure}

\begin{figure}[!h]
    \centering
    \includegraphics[width=0.8\linewidth]{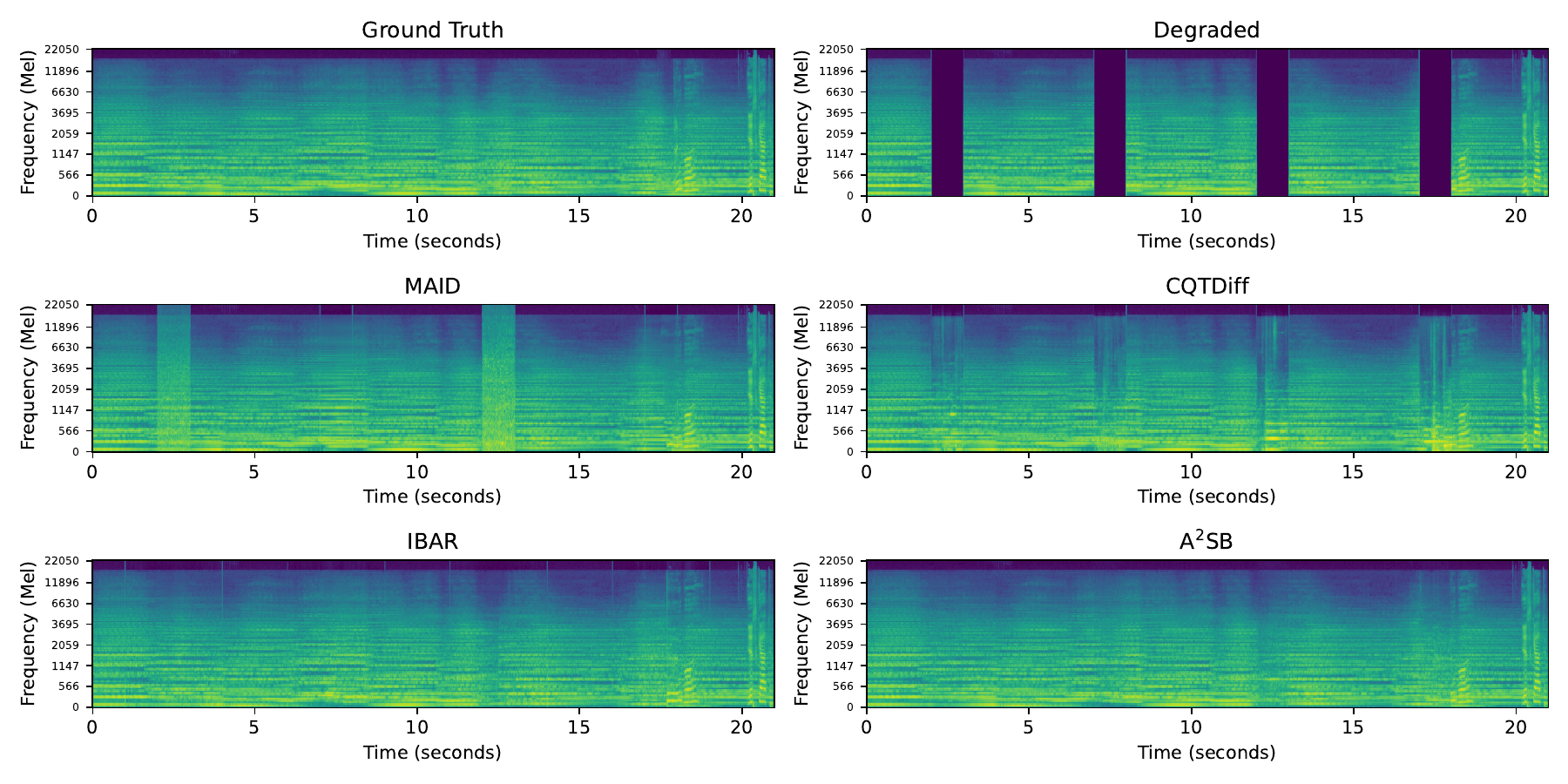}
    \caption{Qualitative comparison between different inpainting methods with inpainting gap = 1 sec.}
    \label{fig: app: inp 2}
\end{figure}

\begin{figure}[!h]
    \centering
    \includegraphics[width=0.8\linewidth]{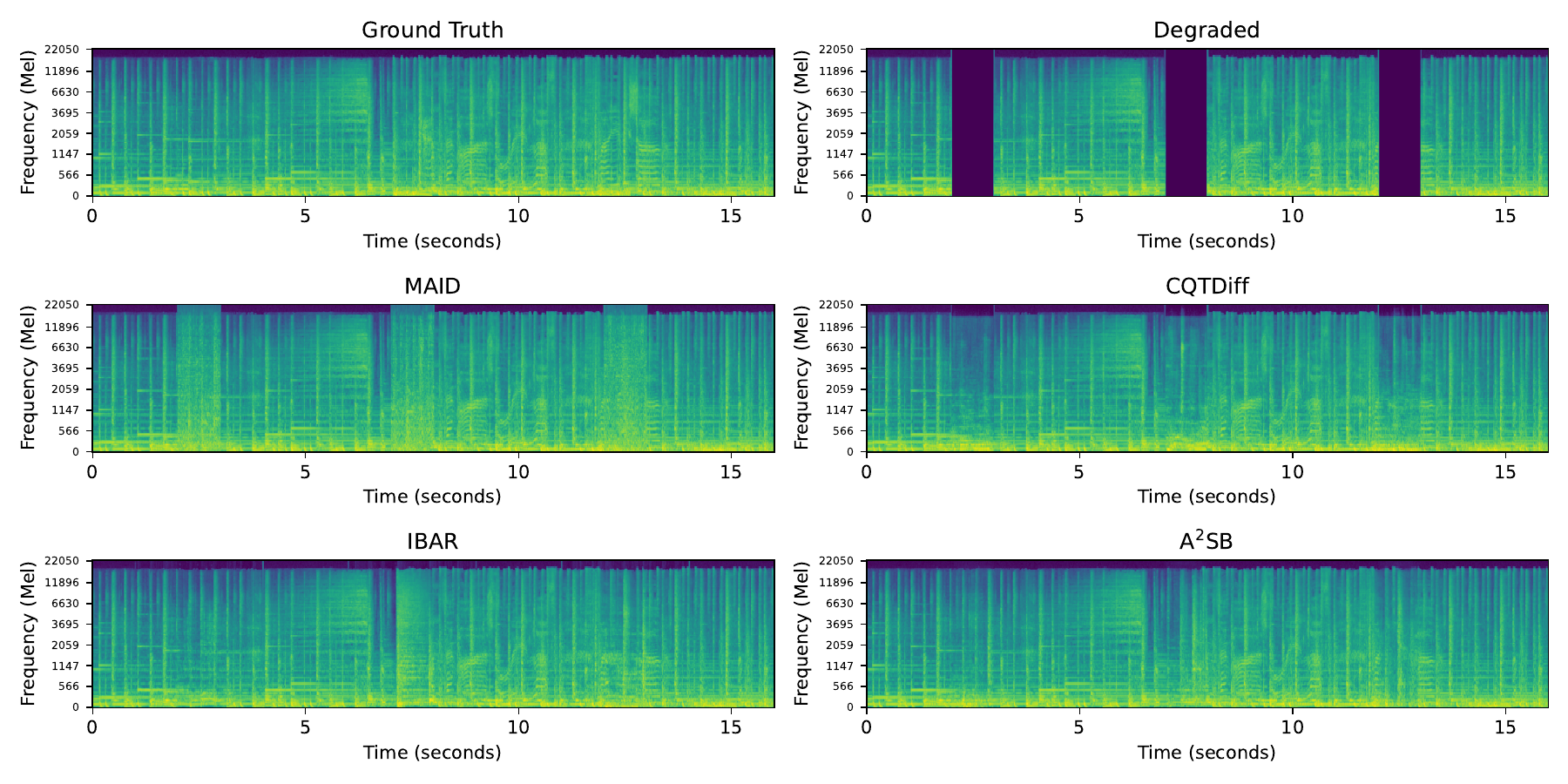}
    \caption{Qualitative comparison between different inpainting methods with inpainting gap = 1 sec.}
    \label{fig: app: inp 3}
\end{figure}

\newpage

\begin{figure}[!h]
    \centering
    \includegraphics[width=0.8\linewidth]{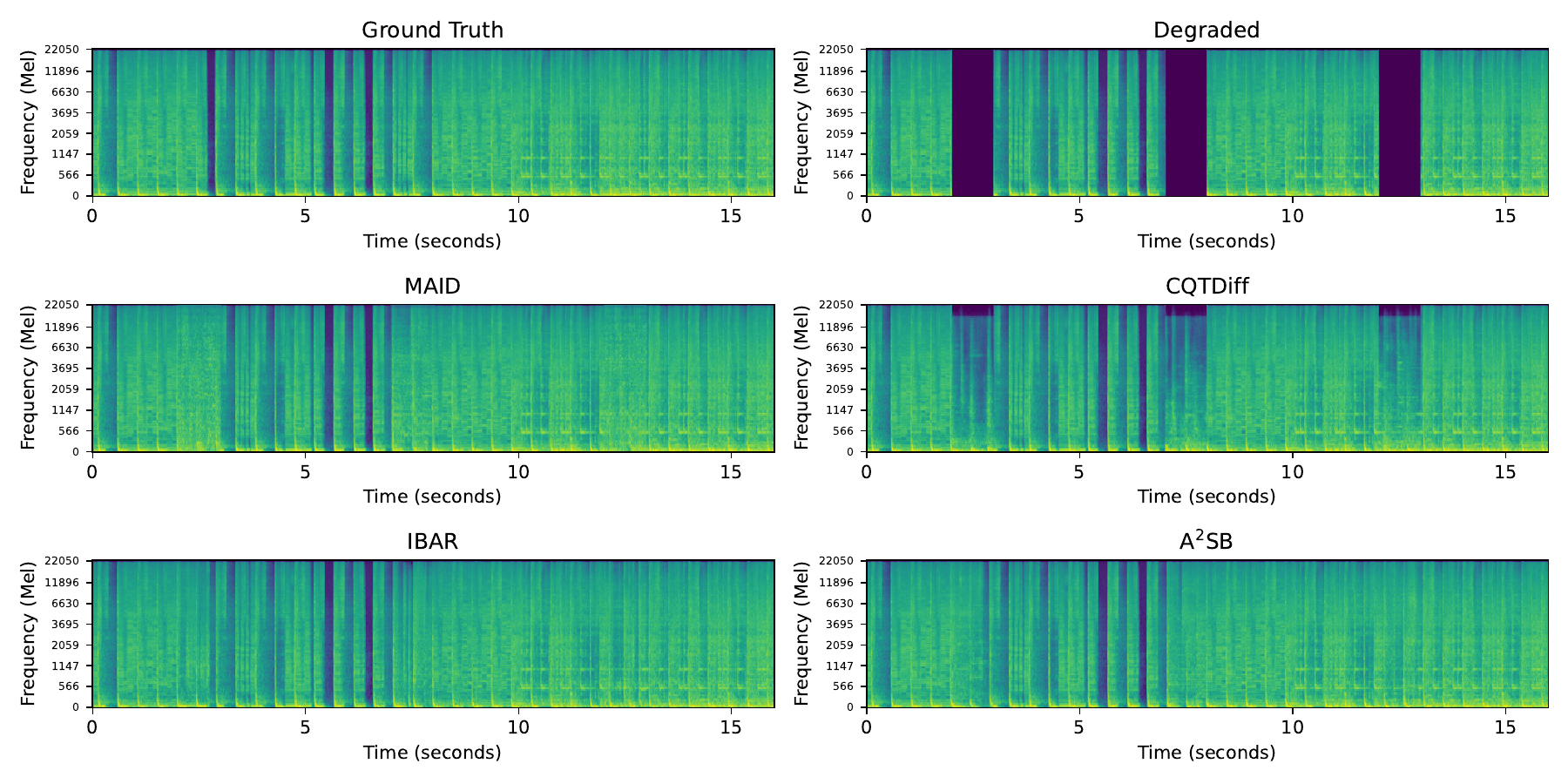}
    \caption{Qualitative comparison between different inpainting methods with inpainting gap = 1 sec.}
    \label{fig: app: inp 4}
\end{figure}

\begin{figure}[!h]
    \centering
    \includegraphics[width=0.8\linewidth]{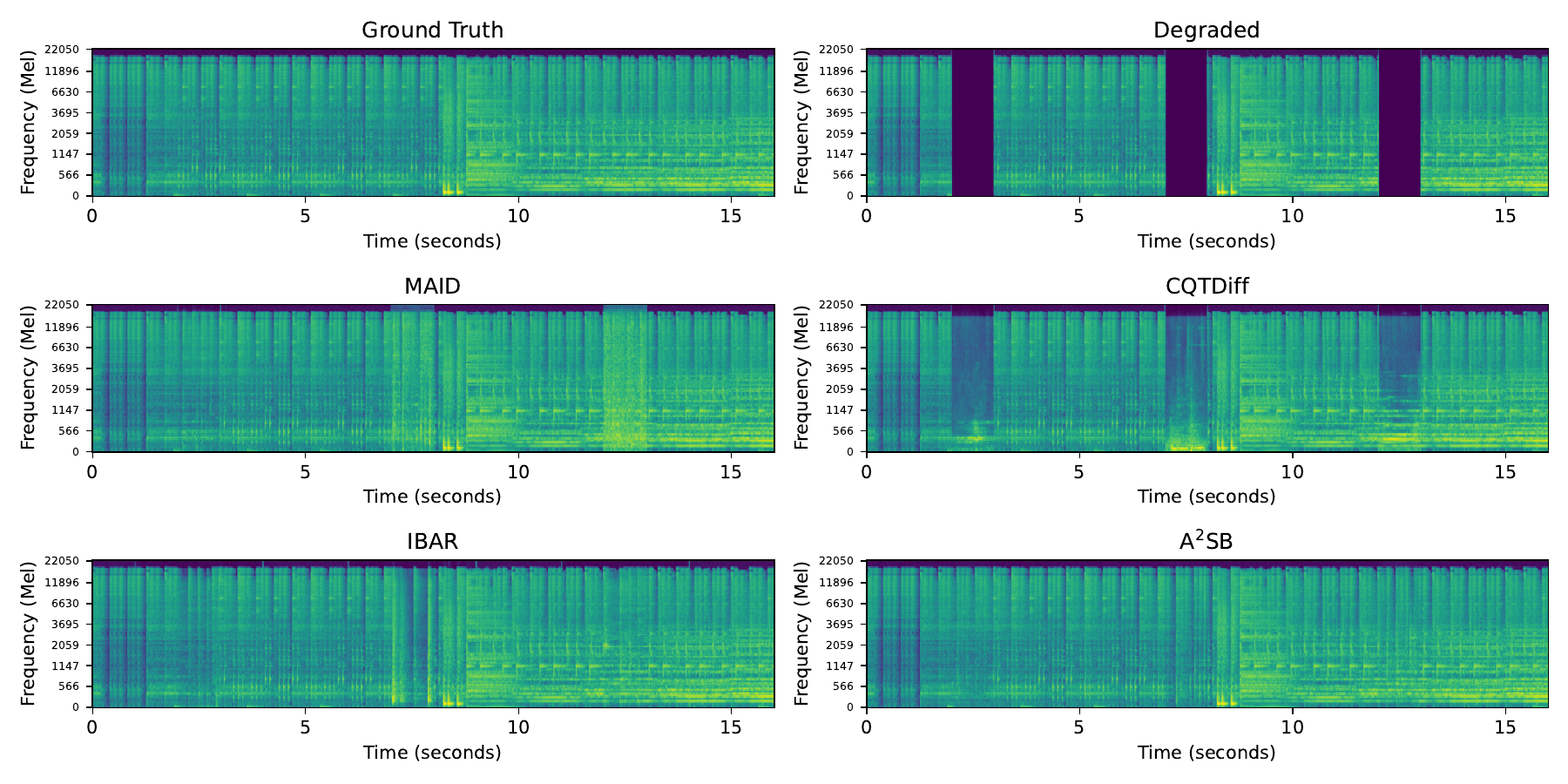}
    \caption{Qualitative comparison between different inpainting methods with inpainting gap = 1 sec.}
    \label{fig: app: inp 5}
\end{figure}

\begin{figure}[!h]
    \centering
    \includegraphics[width=0.8\linewidth]{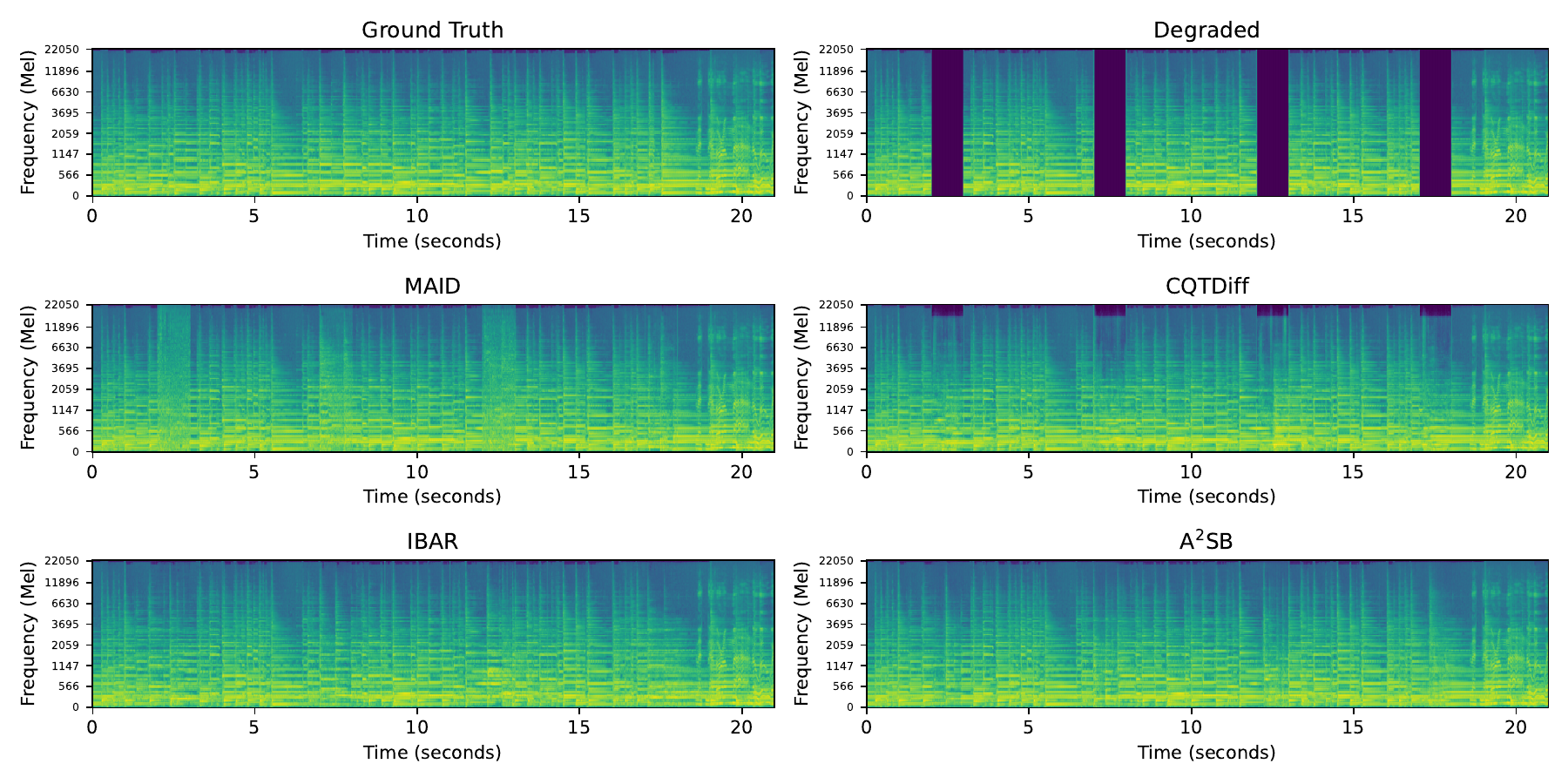}
    \caption{Qualitative comparison between different inpainting methods with inpainting gap = 1 sec.}
    \label{fig: app: inp 6}
\end{figure}

\end{document}